\documentclass[twocolumn]{aastex61}  

\usepackage{mathtools}
\usepackage{amsmath}
\usepackage{graphicx}
\usepackage[all]{hypcap}								
\usepackage{xcolor}
\usepackage{hyperref}
\usepackage{braket}
\usepackage{fancyhdr}
\usepackage{bm}
\usepackage{natbib}

\hypersetup{linkcolor=cyan,colorlinks=true,filecolor=cyan,citecolor=cyan}

\def\Re{{\rm Re}}

\def\tocm{$21\,\textrm{cm}$\ }

\begin{document}

\title{Mitigating Internal Instrument Coupling for 21 cm Cosmology I:\\
Temporal and Spectral Modeling in Simulations}
\shorttitle{Mitigating Internal Instrument Coupling for 21 cm Cosmology I}
\shortauthors{Kern et al.}

\correspondingauthor{Nicholas Kern}
\email{nkern@berkeley.edu}
\author{Nicholas S. Kern}
\affiliation{Department of Astronomy, University of California, Berkeley, CA}

\author{Aaron R. Parsons}
\affiliation{Department of Astronomy, University of California, Berkeley, CA}

\author{Joshua S. Dillon}
\affiliation{Department of Astronomy, University of California, Berkeley, CA}
\affiliation{NSF Astronomy and Astrophysics Postdoctoral Fellow}

\author{Adam E. Lanman}
\affiliation{Department of Physics, Brown University, Providence, RI}

\author{Nicolas Fagnoni}
\affiliation{Cavendish Astrophysics, University of Cambridge, Cambridge, UK}

\author{Eloy  de~Lera~Acedo}
\affiliation{Cavendish Astrophysics, University of Cambridge, Cambridge, UK}

\begin{abstract}
We study the behavior of internal signal chain reflections and antenna cross coupling as systematics for 21 cm cosmological surveys.
We outline the mathematics for how these systematics appear in interferometric visibilities and describe their phenomenology.
We then describe techniques for modeling and removing these systematics without attenuating the 21 cm signal in the data.
This has critical implications for low-frequency radio surveys aiming to characterize the \tocm signal from the Epoch of Reionization and Cosmic Dawn, as systematics can cause bright foreground emission to contaminate the EoR window and prohibit a robust detection.
We also quantify the signal loss properties of the systematic modeling algorithms, and show that our techniques demonstrate resistance against EoR signal loss.
In a companion paper, we demonstrate these methods on data from the Hydrogen Epoch of Reionization Array as a proof-of-concept.
\end{abstract}

\defcitealias{Kern2019b}{K19b}


\section{Introduction}
\label{sec:intro}
Highly redshifted \tocm emission from neutral hydrogen in the intergalactic medium (IGM) promises to be a revolutionary tool with which we can study the formation of the first generations of stars and galaxies in the Universe and the impact they had on their large-scale environments \citep{Hogan1979, Madau1997, Tozzi2000}.
The \tocm signal is a sensitive, tomographic probe of the astrophysics and cosmology of Cosmic Dawn, or the era of first luminous structure formation, and the Epoch of Reionization (EoR) where ionizing photons from star and galaxy formation ionized all of the neutral hydrogen in the IGM at $z\sim6$ leftover from recombination \citep[for reviews see][]{Furlanetto2006c, Pritchard2012, Mesinger2016b}.
The Cosmic Dawn and the EoR are critical components of a broader understanding of large-scale structure formation, yet they remain largely unexplored due to the experimental difficulty of systematically probing the Universe at these redshifts across the electromagnetic spectrum.

The prospect of robustly characterizing the \tocm signal from these epochs is daunting, as galactic and extra-galactic foreground emission outshine the fiducial cosmological signal by many orders of magnitude.
Nevertheless, the path towards detecting the \tocm signal from the Cosmic Dawn and EoR has seen tremendous progress over the past decade, as first generation radio interferometric experiments such as the Donald C. Backer Precision Array for Probing the Epoch of Reionization \citep[PAPER;][]{Parsons2014, Jacobs2015, Ali2015}, the Murchison Widefield Array \citep[MWA;][]{Dillon2014, Beardsley2016, Ewall-Wice2016b}, the Low Frequency Array \citep[LOFAR;][]{Patil2017}, and the Giant Metre Wave Radio Telescope \citep[GMRT;][]{Paciga2013} have placed increasingly competitive limits on the \tocm power spectrum, while single-dish experiments may have made a first detection of the global signal \citep{Bowman2018}.
Going forward, second generation interferometric experiments like the Hydrogen Epoch of Reionization Array \citep[HERA;][]{DeBoer2017} and the Square Kilometer Array \citep[SKA;][]{Koopmans2015} are expected to have the raw sensitivity needed to not only detect the \tocm signal but provide a power spectrum characterization across a wide range of redshifts, leading to drastic improvements in our understanding of astrophysical and cosmological parameters that govern large scale structure and star formation at these epochs \citep{Pober2014, Greig2015a, Greig2015b, Liu2016b, Ewall-Wice2016b, Greig2017b, Kern2017}.

Due to the faintness of the cosmological signal, tight control of instrumental systematics is a crucial component of data reduction pipelines.
Indeed, many of the upper-limits provided by first-generation interferometric experiments have already been systematics limited.
As second generation experiments are constructed and get closer to making a first detection of the \tocm power spectrum, our ability to model and remove systematics to high dynamic range will be of utmost importance in maximizing the scientific impact of future \tocm datasets.
Systematic contamination can be generated in a variety of ways, such as calibration errors, ionospheric Faraday rotation, primary beam ellipticity, analog signal chain imperfections (such as impedance mismatches), and others.
Generally, their end-result is to distort the strong foreground signal in the data, thus making it harder to separate it from the underlying \tocm signal.

In this work, we focus specifically on a class of systematics that we refer to as \emph{internal instrument coupling}, which we further break down into two sub-categories: 1) signal chain reflections or coupling within an antenna signal chain and 2) antenna cross coupling, or coupling between antenna signal chains.
Signal chain reflections are generated by impedance mismatches between transmitting surfaces in the analog signal chain when the signal is carried as a voltage.
Their effect is to generate a copy of the foreground signal in a region of $k$ space nominally occupied only by the EoR \tocm signal and thermal noise.
Cross coupling, on the other hand, can occur from a variety of mechanisms, but common sources for radio surveys are stray capacitance between parallel wires or circuit lines in the signal chain (i.e. capacitive crosstalk) and reflections between antennas in the field (i.e. mutual coupling).
Cross coupling produces a spurious phase-stable term in the data across time that can occupy a wide range of $k$ modes depending on its origin.
These systematics are of critical concern for low-frequency radio surveys \citep{Parsons2012b, Zheng2014, Chaudhari2017}, and have proven to be a partially limiting factor in previous \tocm interferometric analyses \citep{Beardsley2016, Ewall-Wice2016b}.
We use HERA sky and systematic simulations to study the temporal and spectral behavior of internal coupling systematics in the context of HERA data, and proposes techniques for modeling and removing them from the interferometric data products.
We furthermore quantify the signal loss properties of our algorithms with ensemble subtraction trials containing simulated EoR observations.
In this work we do not simulate the effects of thermal noise, as we aim to probe the inherent algorithmic performance to very high dynamic range, however, we discuss the anticipated impact of thermal noise on our algorithms in \autoref{sec:modeling}.

Recently, \citet{Tauscher2018} studied systematic removal for \tocm survey data using a simulation-based, forward-model approach for separating systematics from the EoR signal.
This seems to work well when the systematic parameter space is fairly limited and the systematic itself well-understood: in their case they consider only beam-weighted foregrounds as a systematic, with a few parameters governing its spatial and spectral dependence.
Our work takes a semi-empirical approach to modeling instrument systematics due to 1) the highly variant nature of the systematics observed in the real data, and 2) because we require suppression to high dynamic range, which is often more easily achieved with flexible empirical methods. 
The consequence of using semi-empirical systematic models is that they can be overfit and induce loss of EoR signal, which we test for in this work.

The structure of this paper is as follows.
In \S2 we provide a mathematical overview of how the two systematics described in this paper corrupt interferometric data products, and make predictions for their spectral and temporal behavior.
In \S3 we introduce algorithms for modeling and removing these systematics from the data, and then use simulated HERA data with and without systematics to test the performance of our techniques.
In \S4 we validate our algorithms by quantifying their signal loss properties, ensuring they are not lossy to the desired EoR sky signal.
Lastly, in \S5 we summarize our results.
In a companion paper, \citet[][]{Kern2019b}, we demonstrate the performance of these techniques on real data from HERA Phase I instrument.
Many of the assumptions made in this work about the functional form of the simulated systematics come from the actual observed systematics outlined in that paper.

\begin{figure}
\centering
\label{fig:sigchain}
\includegraphics[scale=0.35]{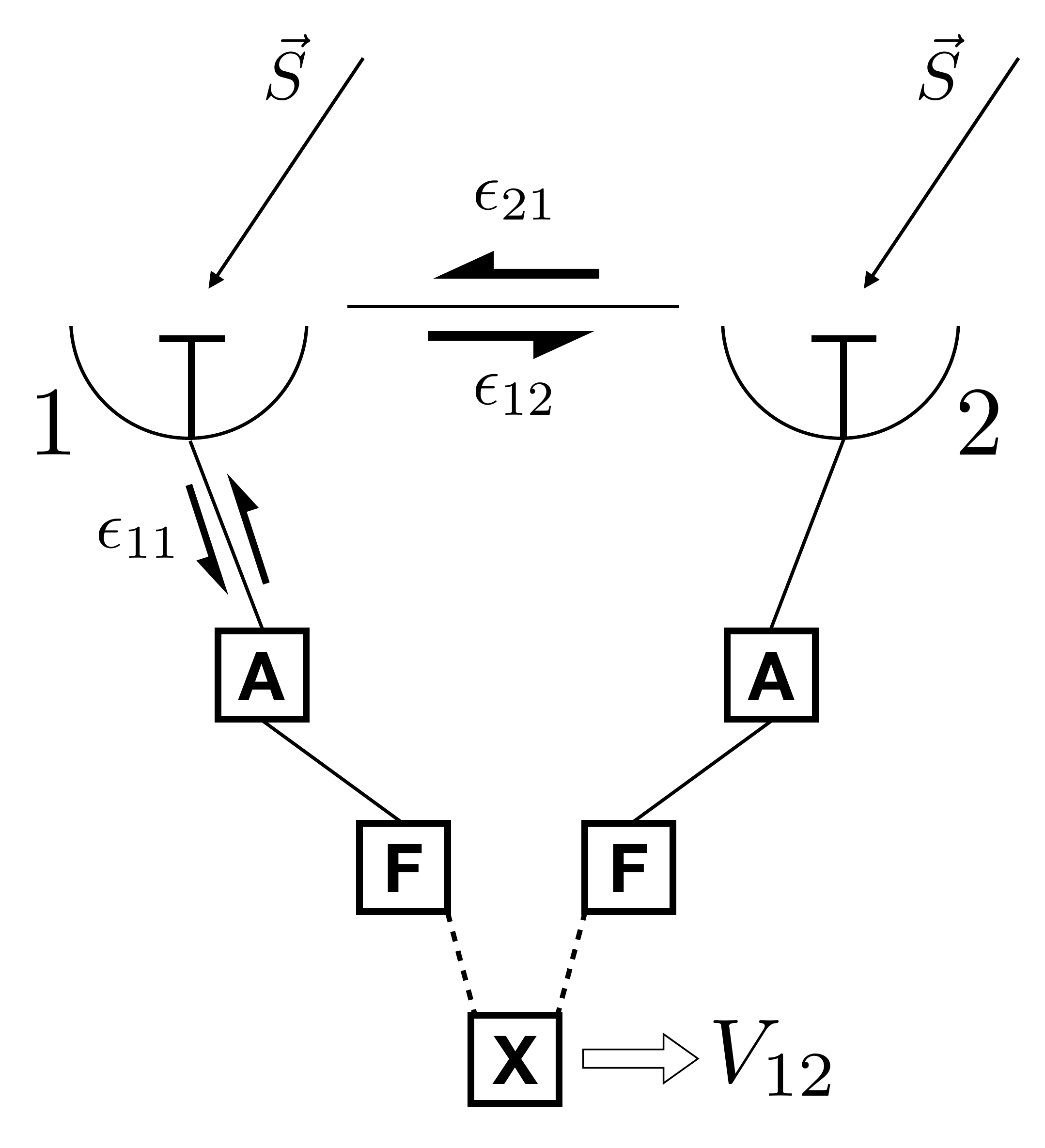}
\caption{A schematic of two HERA signal chains, 1 \& 2, with possible sources of systematics demarcated.
Sky signal ($\vec{\mathcal{S}}$) enters each antenna's feed, is converted into a voltage and travels down their signal chains where it is first processed at a node housing an amplifier ($\mathbf{A}$). It is then directed to an engine that digitizes and Fourier transforms the signal ($\mathbf{F}$) and sent to the correlator ($\mathbf{X}$), which produces the visibility $V_{12}$. A possible cable reflection in antenna 1's signal chain is marked as $\epsilon_{11}$, traversing up and down the cable connecting the feed to the node, and possible cross-coupilng is marked as $\epsilon_{12}$, where radiation is reflected off of antenna 2 and into antenna 1, or vice versa. Dashed lines indicate a signal pathway after digitization, where internal instrument coupling is no longer a major concern.}
\end{figure}

\section{Mathematical Overview}
\label{sec:prelim}
In this section, we describe how signal chain reflections and antenna cross couplings appear in interferometric data products.
To begin, we start with a the two-element interferometer \citep{Hamaker1996, Smirnov2011},
consisting of two antennas, 1 and 2, whose feeds measure an incident electric field and convert it into a voltage.
In \autoref{fig:sigchain}, we show a schematic of the HERA analogue system and mark possible sources of internal instrument coupling.
These signals travel from the feeds through each antenna's signal chain to the correlator,
and along the way are amplified, digitized, channelized, and Fourier transformed into the frequency domain.
The correlator then cross multiplies voltage spectra to form the fundamental interferometric data product: the 
cross-correlation visibility, $V_{12}$, between antenna 1 and 2, written as
\begin{align}
\label{eq:ME}
V_{12}(\nu, t) = v_1(\nu, t) v_2^\ast(\nu, t).
\end{align}
The correlator can also produce the auto-correlation visibility by correlating an antenna voltage with itself (e.g. $V_{11}$).
Here we have chosen to define the visibility as the product of two antenna voltage spectra, rather than the correlation of voltage time streams: although the two are equivalent given the convolution theorem, the former will prove to be an easier basis when working with reflections.
In addition, we have been explicit about the frequency and time dependence of each antenna's voltage spectrum $v$ and, by extension, the complex visibility $V$, although we drop these throughout the text for brevity.
We have also dropped the time averaging done by any real correlator, which is done for brevity and does not alter our results in this section.
While we could have cast the visibility equation (\autoref{eq:ME}) in matrix form \citep[e.g.][]{Hamaker1996, Smirnov2011}, we find it easier to understand the impact that the specific sytematics we discuss in this paper have on the resultant data products using a simpler, algebraic form for the visibility equation.

\begin{figure*}
\centering
\label{fig:sim_waterfalls}
\includegraphics[scale=0.5]{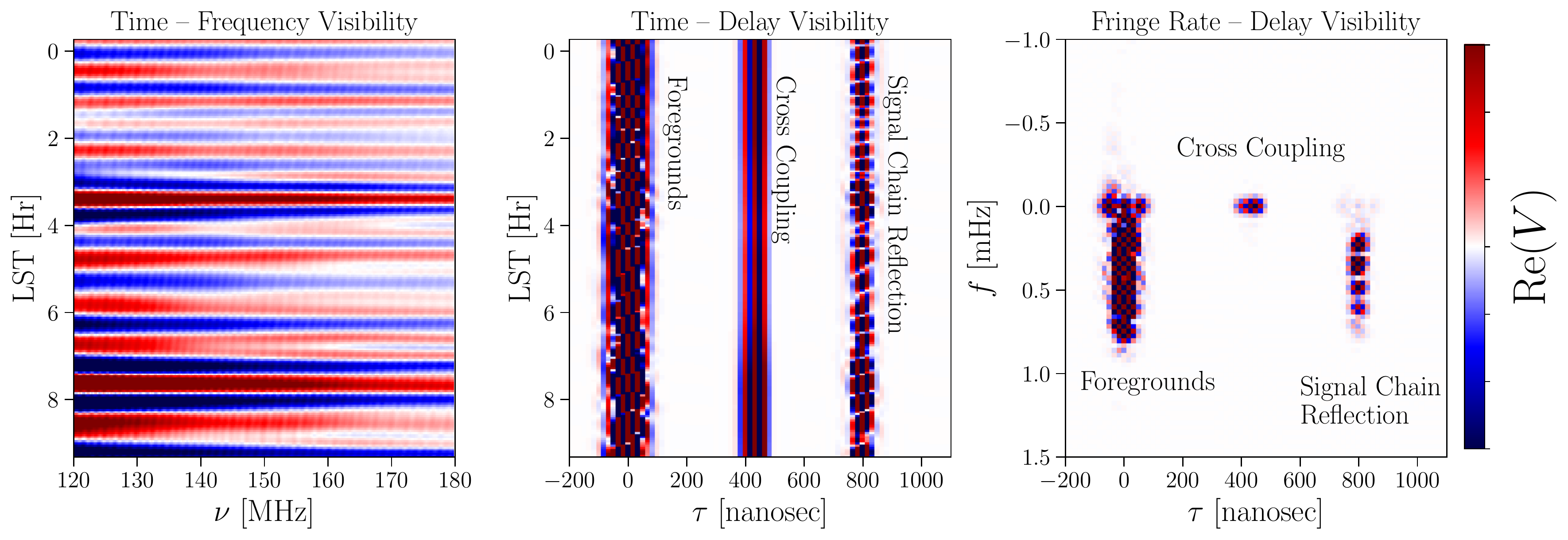}
\caption{The real component of a simulated cross-correlation visibility with foregrounds, a signal chain reflection inserted at $\tau=800$ ns and a cross coupling term inserted at $\tau=400$ ns, plotted in dimensionless units for visual clarity.
{\bfseries Left:} Visibility in time and frequency space. {\bfseries Center:} Visibility in time and delay space. {\bfseries Right:} Visibility in fringe-rate and delay space. Different components of the visibility--in particular systematics--are usually better separated in delay and fringe-rate space than in the original time and frequency space.}
\end{figure*}

The correlator outputs time-ordered visibilities as a function of local sidereal time (LST; denoted as $t$) and frequency (denoted as $\nu$).
When Fourier transforming the data across the frequency axis, we put the data into a temporal domain.
To separate this from the original time domain, we refer to the Fourier dual of frequency as the \emph{delay domain} (denoted as $\tau$).
Similarly, the Fourier transform of our data across time puts the data into a spectral domain, which we refer to as the \emph{fringe-rate domain} (denoted as $f$), using similar notation as \citet{Parsons2016}.
In the absence of explicit markers, we will use $V$ to mean the visibility in time and frequency space, and use $\widetilde{V}$ to mean the visibility in one or both of the Fourier domains: which should be clear based on context, otherwise we will use explicit notation..

Different components of the visibility are generally more localized in Fourier space.
Foreground signal, for example, is intrinsically spectrally smooth and will therefore occupy low delay modes, whereas a fiducial EoR model, being non-spectrally smooth, occupies low and high delay modes.
\autoref{fig:sim_waterfalls} shows a simulated foreground + systematic visibility in real and Fourier space, demonstrating how systematics are usually better separated in Fourier spaces (center \& right panel).
We present the figure here to guide the reader's intuition about the phenomenology of the systematics in real and Fourier space while we discuss their mathematical form below.
Note that the parameters of the systematics as simulated in \autoref{fig:sim_waterfalls}, for example the delays they show up at, have been chosen merely for visual clarity, and are not necessarily the systematic parameters seen in actual data.
We describe the simulations used throughout this work in \autoref{sec:vis_sims}.

\subsection{Describing Signal Chain Reflections}
A reflection in the signal chain of an antenna inserts a copy of the original signal with a time lag.
An example of this is a reflection at the end of an analogue cable, where the signal travels back up and cable, reflects again at the start of the cable and travels back down and is transmitted through the system along with the original signal.
The time lag, or delay, the reflected signal has acquired is two times the cable length divided by the speed of light in the cable.
The reflected signal also acquires an amplitude suppression meaning it is generally only a fraction of the input signal, but even a small fraction of the foreground signal in the data can dwarf the expected EoR signal and therefore needs to be accounted for.
If $v_1$ is antenna 1's voltage spectrum without a signal chain reflection, then the presence of a reflection can be encapsulated as
\begin{align}
\label{eq:reflection}
v_1^\prime(\nu, t) = v_1(\nu, t) + \epsilon_{11}(\nu) v_1(\nu, t)
\end{align}
where $v_1^\prime$ is the voltage spectrum of antenna 1 with the reflection component, and $\epsilon$ is a coupling coefficient describing the reflection in antenna 1's signal chain (denoted as $11$ because it is coupling the signal with itself).
The coupling coefficient can be broken into three constituent parameters as
\begin{align}
\label{eq:epsilon}
\epsilon_{11}(\nu) = A_{11}e^{2\pi i\tau_{11}\nu + i\phi_{11}},
\end{align}
where $A$ is the amplitude, $\tau$ is the delay offset (the total time it takes to be reflected) and $\phi$ is the
phase offset the reflected signal may have acquired relative to the original signal.
In \autoref{eq:reflection} we have assumed time and frequency stability of the reflection parameters, although in practice these parameters will have some variation with time and frequency.

If we insert the corrupted voltage spectra into the visibility equation (\autoref{eq:ME}), we get
\begin{align}
\label{eq:cross_corr_sig_ref}
V_{12}^\prime = v_1v_2^\ast + \epsilon_{11}v_1v_2^\ast + v_1\epsilon_{22}^\ast v_2^\ast + \epsilon_{11}v_1\epsilon_{22}^\ast v_2^\ast.
\end{align}
We can see that in addition to the original cross-correlation term ($v_1v_2^\ast$) we now also have copies of it at positive and negative delay offsets that are suppressed in amplitude by a factor of $A_{11}$ and $A_{22}$, respectively. The time-behavior of a reflection mimics that of the original data, in that it shows the same temporal oscillation (i.e. fringing) as the foregrounds, and thus also appears at the same fringe-rate modes as the foregrounds (e.g. right panel of \autoref{fig:sim_waterfalls}).
The conjugation of $\epsilon_{22}$ means that the reflected signal from antenna 2 appears at negative delays in $V_{12}$, while the reflected signal from antenna 1 appears at positive delays.


The resultant auto-correlation visibility can also be computed, and is given by
\begin{align}
\label{eq:auto_corr_sig_ref}
V_{11}^\prime = v_1v_1^\ast + \epsilon_{11}v_1v_1^\ast +  v_1\epsilon_{11}^\ast v_1^\ast + |\epsilon_{11}|^2v_1v_1^\ast,
\end{align}
where we see that the first order reflections show up at $\pm\tau_{11}$, while the second-order reflection appears at $\tau=0$ ns due to the conjugation of the coupling coefficient with itself.
If we assume that the first order reflections are at sufficiently high delay and neglect the second-order term, we can approximate the visibility at $\tau=0$ ns as $\widetilde{V}_{11}^\prime(\tau=0) \approx (v_1v_1^\ast)(\tau=0)$.
Similarly, near the reflection delay of $\tau_{11}$ the auto-correlation visibility simplifies to
\begin{align}
\widetilde{V}_{11}^\prime(\tau=\tau_{11}) \approx \epsilon_{11}v_1v_1^\ast.
\end{align}
This means that one can estimate the reflection coefficient amplitude in delay space as
\begin{align}
\label{eq:ref_amp}
A_{11} = \frac{\left|\widetilde{V}_{11}^\prime(\tau=\pm\tau_{11})\right|}{\left|\widetilde{V}_{11}^\prime(\tau=0)\right|},
\end{align}
which will be useful when modeling reflection systematics.

If one can estimate their parameters from the data, reflections can be removed via standard (direction-independent) antenna based calibration.
In this paradigm, the raw voltage spectrum of antenna 1 corrupted by the instrument  is related to its true value as
\begin{align}
v^{\rm raw}_1 = v_1 g_1,
\end{align}
which when inserted into the visibility equation yields the standard antenna based calibration equation,
\begin{align}
V^{\rm raw}_{12} = V_{12}g_1g_2^\ast = \langle v_1 v_2^\ast\rangle g_1g_2^\ast.
\end{align}
The $g$ term is called the antenna gain, and accounts for amplitude and phase errors introduced by the various stages of the signal chain from the feed all the way to the correlator.
Note that this form of the calibration equation does not account for polarization leakage induced by cross-feed coupling, which is generally a higher order correction \citep{Hamaker1996, Sault1996}.
By re-arranging \autoref{eq:reflection} as
\begin{align}
\label{eq:ref_gains}
v_1^\prime = v_1(1 + \epsilon_{11}) = v_1g_1,
\end{align}
we can see that signal chain reflections can be completely encompassed in this gain term, and hence corrected for by dividing the corrupted data by a gain constructed from an estimate of the reflection coefficient.

\subsection{Describing Antenna Cross Coupling}
We now turn our attention to another systematic we refer to as antenna cross coupling, which acts to couple one antenna's voltage stream with another antenna's voltage stream before reaching the correlation stage.
Note that our model for cross coupling is different than ``capacitive crosstalk'' created by the electric field of two parallel signal chains interacting with each other within cabling, receivers, and analogue-to-digital conversion (ADC) units, which is a common systematic for radio interferometers \citep{Parsons2009, Zheng2014, Ali2015, Patil2017, Cheng2018}.
There are well established hardware solutions for suppressing crosstalk, such as phase switching \citep{Chaudhari2017}.
However, if residual crosstalk remains, or if phase switching is not implemented in the system, we need to model and remove it for robust EoR measurements.

Our cross cross systematic model simply states that, before correlation, one antenna's voltage is added to another antenna's voltage with a coupling coefficient that determines the amplitude and relative delay with which the voltage is added.
For the purposes of our description, we additionally assume that this coupling coefficient can be decomposed into the same three parameters as before, technically making it a form of reflection systematic.\footnote{While we adopt this assumption in this section to make the algebra simpler, the algorithm we present in \autoref{sec:modeling} is more general and does not rely on this assumption.}
While this model may indeed be capable of describing certain some forms of capacitive crosstalk, we do not expect all forms of capacitive crosstalk to necessarily fall within the bounds of these assumptions.

To write down how this affects the interferometric visibility, we can start by writing the corrupted antenna voltages as
\begin{align}
\label{eq:cross_ant_volt}
v_1^\prime &= v_1 + \epsilon_{21}v_2 \nonumber \\
v_2^\prime &= v_2 + \epsilon_{12}v_1,
\end{align}
where $\epsilon_{21}$ describes the voltage coupling of antenna 2 into antenna 1 and vice versa for $\epsilon_{12}$.
Substituting these equations into \autoref{eq:ME}, we get
\begin{align}
\label{eq:coupling_cross_corr}
V_{12}^\prime = v_1v_2^\ast +  v_1\epsilon_{12}^\ast v_1^\ast +  \epsilon_{21}v_2v_2^\ast +  \epsilon_{21}v_2\epsilon_{12}^\ast v_1^\ast.
\end{align}
We can see that the cross-correlation visibility now contains the auto-correlation visibility terms $v_1v_1^\ast$ and $v_2v_2^\ast$ at the first-order level, which are purely real quantities and thus have identically zero phase.
In the complex plane, the cross-correlation term $v_1v_2^\ast$ winds around the origin as a function of time because its phase varies temporally.
Because the auto-correlation has no phase, the act of the cross coupling terms is to introduce an additive bias to the data with an arbitrary phase set by the coupling coefficient itself.
Assuming the coupling coefficient is slowly variable (if not completely stable), the first order systematic terms in \autoref{eq:coupling_cross_corr} only change in amplitude over time set by the natural variation in the amplitude of the auto-correlation, (e.g. $v_1v_1^\ast$).
This variation is generally fairly slow on timescales of a beam crossing, which for HERA is roughly 1 hour.
This leads us to two critical insights about the behavior of the cross coupling terms: 1) their time variability is slow, thus occupying low-fringe rate modes (e.g. see right of \autoref{fig:sim_waterfalls}) and 2) they have a time-stable phase determined solely by the phase of the coupling coefficient.

In a more generalized case, the cross coupling between antenna 1 and 2 may have an angular dependence on the sky.
Take for example the case of mutual coupling (or feed-to-feed reflections), where part of the radiation incident on antenna 1's feed is reflected and received by antenna 2's feed.
This behavior will be highly angular dependent due to the non-trivial electromagnetic properties of the feed itself.
Nonetheless, we can reason that the systematic phenomenology will be similar to as before.
We can think of this angular dependence as a windowing function on the primary beam of the underlying auto-correlation, meaning that
the first-order terms in \autoref{eq:coupling_cross_corr} will be proportional to only a fraction of $v_1v_1^\ast$, such that they have a smaller amplitude.
It may also mean that these terms will have a slightly faster time dependence in the data, as the ``effective beam'' created by the angular windowing function is smaller on the sky than the total primary beam, and thus leads to a faster ``effective beam crossing time.''

We can also compute the effects of cross coupling on the measured auto-correlation visibility, $V_{11}$, which yields
\begin{align}
V_{11}^\prime = v_1v_1^\ast + v_1\epsilon_{21}^\ast v_2^\ast + \epsilon_{21}v_2v_1^\ast + |\epsilon_{21}|^2v_2v_2^\ast.
\end{align}
In this case, we find that the cross-correlation is inserted into the measured auto-correlation at the first order level with a delay offset of $\tau_{21}$.
These terms are likely many order of magnitudes below the peak auto-correlation visibility amplitude, given that the cross-correlation visibilities are generally a few orders of magnitude below the auto-correlation inherently, which is further compounded by the amplitude suppression from $\epsilon_{21}$.

We can see simply from \autoref{eq:coupling_cross_corr} that the corruption of $V_{12}^\prime$ by cross coupling \emph{cannot} be factorized into antenna based gains, based simply on the presence of the $\epsilon_{12}$-like terms, which are baseline-dependent.
Removal of cross coupling terms in the data must therefore be done on a per-baseline basis by constructing a model of the systematic in each visibility and then subtracting it.

\subsection{Summary}
To summarize, reflections along a single antenna's signal chain produces a duplicate of the signal with suppressed amplitude and some delay offset. 
This is true for both the cross and auto-correlation visibility products.
Example mechanisms include cable reflections and dish-to-feed reflections within the confines of a single antenna.
Reflections in the cross-correlation visibility have the same time structure as the un-reflected visibility, meaning reflected foreground signal occupies the same fringe-rate modes as un-reflected foreground signal, but is shifted to high delays (e.g. \autoref{fig:sim_waterfalls}).
Reflections can be removed from the raw data by creating a model of the reflections and incorporating them into the per-antenna calibration gains.

Another systematic we describe is created by antenna-to-antenna cross coupling, which mixes the voltage signals between the antennas.
This has the effect of introducing a copy of the auto-correlation visibility into the measured cross-correlation visibility at positive and negative delay offsets, and similarly introduces copies of the cross-correlation visibility into the measured auto-correlation visibility.
In the measured cross-correlation visibility, the first-order coupling terms are slowly time variable and occupy low fringe-rate modes centered at $f=0$ Hz.
Cross coupling terms cannot be removed via antenna based calibration, and must be modeled and subtracted at the per-baseline level.


\section{Systematic Modeling}
\label{sec:modeling}

Next we discuss our approach for modeling reflection and cross coupling systematics in the data.
We use sky and instrument simulations to generate mock visibilities of diffuse foregrounds and systematics, which we use to test our algorithms and provide benchmarks on their performance.
More details on the construction of the simulated data products used in this work can be found in \autoref{sec:vis_sims}.
The fiducial parameters of our systematic simulations are chosen to roughly reflect the behavior of systematics seen in HERA Phase I data \citep{Kern2019b}.
Systematics in the HERA Phase I system can be found at variable amplitudes and delays in the data depending on the baseline or antenna at hand, but are generally seen at an amplitude of $\sim3\times10^{-3}$ times the peak foreground amplitude at $\tau=0$ ns and at delays spanning 200 -- 1500 ns.
For these simulations we do not include instrumental thermal noise so that we can test the underlying performance of our algorithms to high dynamic range.

\subsection{Modeling Signal Chain Reflections}

\begin{figure*}
\centering
\label{fig:auto_fitting}
\includegraphics[scale=0.55]{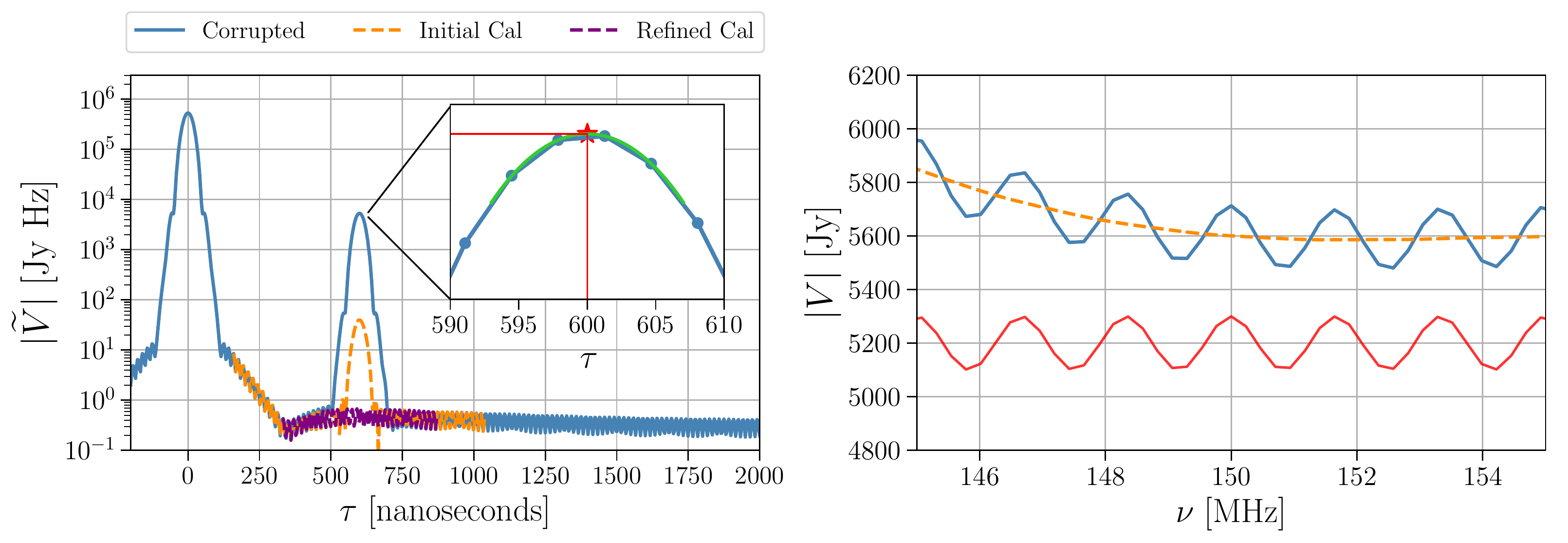}
\caption{Reflection modeling and removal on a simulated auto-correlation visibility.
{\bfseries Left}: Foreground-only auto-correlation in delay space with a simulated cable reflection at 600 ns (blue).
Dashed orange shows the visibility after initial reflection calibration, demonstrating roughly two orders of magnitude of suppression.
The inset highlights the reflection bump, showing a spectral fit via quadratic interpolation (green curve) to
achieve more precise estimates of the reflection delay and amplitude (red star). The calibration is then refined using an iterative technique until the reflection bump is minimized (purple).
{\bfseries Right}: Simulated visibility with reflection in frequency space (blue), a scaled version of the fitted reflection coefficient (red) highlighting its phase coherence with the reflection ripple in the data, and the visibility after initial calibration (dashed orange).}
\end{figure*}

Modeling signal chain reflections can in theory be done simultaneously with standard gain calibration because reflections factor as an antenna-based effect (see \autoref{sec:prelim}).
However, there are reasons why we might be wary of using standard calibration techniques for deriving reflection parameters.
Principally, standard bandpass calibration typically operates on the $\sim N_{\rm ant}^2$ number of cross-correlation visibilities, and generally allows each frequency channel's gain to be solved independently from other channels.
Frequency dependent calibration errors will therefore set a fundamental floor to the precision with which reflections can be calibrated via standard antenna-based bandpass techniques \citep{Barry2016, Ewall-Wice2017, Orosz2018}.
Furthermore, the dynamic range of signal to noise is considerably higher in the auto-correlation than in the cross-correlation visibility, as it is a measurement of the total power received by an antenna: in many cases a signal chain reflection cannot even be seen above the noise floor in a cross-correlation visibility but is highly apparent in the auto-correlation visibility.
This latter point is important, because it implies that reflection parameters estimated from the auto-correlation visibilities will have a signal-to-noise ratio (SNR) that is drastically higher than the SNR of the cross-correlation visibilities, meaning that, when calibrating out systematics in the cross-correlation visibilities, the SNR of the derived reflection parameters will never be a limiting factor.
Of course other real-world factors can limit the precision of the derived reflection parameters such as non-trivial frequency evolution, which is discussed in more detail in the context of HERA in \citet{Kern2019b}.

Reflection parameters must be estimated to high precision in order to get even a modest suppression of their systematic power in the visibilities.
For example, \citet{Ewall-Wice2016b} employed a reflection fitting algorithm on MWA auto-correlation visibilities by fitting sinusoids in the frequency domain and was able to suppress reflection systematics by a couple orders of magnitude in the power spectrum, although their end result band powers were still systematic limited at some $k$ modes.
Similarly, \citet{Beardsley2016} explore reflection calibration on MWA data as an extension to their restricted polynomial gain calibration scheme and also achieve a couple of orders of magnitude of suppression in the 2D power spectrum, although their more deeply integrated power spectra show its re-emergence.

The algorithm we present here also operates on the auto-correlation visibility, but we choose to model the reflection in the delay domain rather than the frequency domain.
Recall from \autoref{sec:prelim} that in the auto-correlation visibility, $V_{11}$, a signal chain reflection appears as a shifted copy of the original visibility at symmetric positive and negative delay offsets.
The Fourier transformed auto-correlation visibility, $|\widetilde{V}_{11}|$, is intrinsically quite peaky in delay space, meaning that a reflection is essentially a narrow spike appearing at its corresponding reflection delay.
The reflection amplitude is then estimated via \autoref{eq:ref_amp}, and the reflection phase is estimated by transforming back to frequency space and aligning reflection templates while varying their phase until a squared error metric is minimized.
This yields an initial estimate of the reflection parameters, but it needs to be refined in order to suppress the systematic to high dynamic range.
In order to refine our parameter estimates, we setup a non-linear optimization system that perturbs the initial guesses, applies the calibration to the data in frequency space, transforms to Fourier space, estimates the amplitude of the residual reflection bump and repeats until it is minimized or a stopping threshold is reached.
Solving this with an iterative minimization technique allows us to estimate the reflection parameters with sufficient accuracy to suppress the reflection in our systematic simulations by eight orders of magnitude in the power spectrum.

To summarize, our approach for estimating the reflection parameters from the data takes the following steps:
\begin{itemize}
\item[1.] Zero-pad the auto-correlation in frequency space and apply a windowing function before Fourier transforming to delay space to minimize sidelobe power
\item[2.] Fit for the peak of the reflection bump in $|\widetilde{V}|$ via quadratic interpolation of its nearest neighbors
\item[3.] The estimated reflection delay, $\tau$, is equal to $\tau_{\rm peak}$
\item[4.] The estimated reflection amplitude, $A$, is the ratio $|\widetilde{V}(\tau=\tau_{\rm peak})| / |\widetilde{V}(\tau=0)|$ (\autoref{eq:ref_amp})
\item[5.] Set all modes of $\widetilde{V}$ to zero except the modes nearest $\tau_{\rm peak}$ and Fourier transform back to frequency space to get $V_{\rm filt}$
\item[6.] The estimated reflection phase, $\phi$, is found by minimizing $|V_{\rm filt} - Ae^{2\pi i \nu\tau + i\phi}|^2$ while varying $\phi$ from 0 - 2$\pi$.
\item[7.] Setup a non-linear optimization that perturbs the initial reflection parameter estimates, applies the calibration and transforms to Fourier space until the residual near the original reflection bump is minimized or a stopping threshold is reached.
\end{itemize}
We demonstrate this algorithm on a simulated HERA auto-correlation visibility corrupted by a cable reflection with a (frequency independent) amplitude of $10^{-2}$ at a delay of 600 nanoseconds (\autoref{fig:auto_fitting}).
The natural delay resolution of HERA data is 10 nanoseconds, which is much too coarse to achieve precision estimates of the reflection delay.
Zero-padding the data by a factor of three gets us to a delay resolution of 3 ns, but this is still not precise enough for accurate reflection delay estimates.
By employing quadratic interpolation on the spectral peak, we can recover the input cable delay to roughly $\pm0.1$ ns (left of \autoref{fig:auto_fitting}).
In this idealized, noise-free simulation, the initial reflection calibration estimates the reflection delay to within $\pm0.1$ ns of its true value, and its phase to within $\pm0.01$ radians, yet we only see systematic power suppression of two orders of magnitude in the visibilities.
This is representative of the precision needed to achieve even modest systematic suppression.
With the refined reflection calibration, however, we find we can achieve reflection systematic suppression of up to four orders of magnitude in the visibility, and recover the reflection parameters to within 1 part in $10^6$ of their true value.

Fiducial EoR levels are expected to be roughly $10^{-5}$ times the peak cross-correlation foreground power in the visibility at $k_{\parallel}\sim0.1\ h$ Mpc$^{-1}$, and is generally thought to be even weaker at higher $k_{\parallel}$ \citep{Mesinger2011}.
If reflection systematics have inherent amplitudes of around $10^{-3}$, then a few orders of magnitude of further suppression will push them below expected EoR levels at low $k$.
In practice, non-ideal effects like frequency evolution of the reflection parameters will limit the precision of reflection calibration, which has been observed in real instruments \citep{Ewall-Wice2016b}.
Indeed, inclusion of such effects in our systematic simulation will likely degrade our algorithm performance.
However, if frequency evolution is a limiting factor for reflection calibration, one simple strategy is to split the full bandwidth into multiple sub-bands and perform reflection calibration independently on each of them, with the caveat that non-negligible frequency evolution within each sub-band may need to be mitigated in other ways.
One can also do this more self-consistently by estimating the reflection parameters and their frequency dependence across the full band, however, we defer this to future work.

\subsection{Modeling Cross Coupling}
\label{sec:modeling_cross_coupling}
Antenna cross coupling systematics are a baseline-dependent effect and as such must be modeled and subtracted for each cross-correlation visibility independently.
In other works, cross coupling has been modeled as a phase-stable term in the data that can be removed by applying a finite impulse response (FIR) high-pass filter to the data \citep{Parsons2010, Ali2015, Kolopanis2019}.
The algorithm described in this work is conceptually quite similar in that we take advantage of cross coupling's slow time variability to model it, but is different in its methodology.
A comparison of cross coupling subtraction techniques is done in \autoref{sec:fr_filtering}.
Note that the method presented here does not assume that the instrumental bandpass has been calibrated out: these two steps are in principle interchangeable.
In practice, it helps if at least the antenna cable delays are calibrated out such that the main foreground lobe shows up at the expected delays of $\tau\approx0$ ns, but again this is not strictly necessary.


\begin{figure*}
\centering
\label{fig:sim_xtalk_sub}
\includegraphics[scale=0.45]{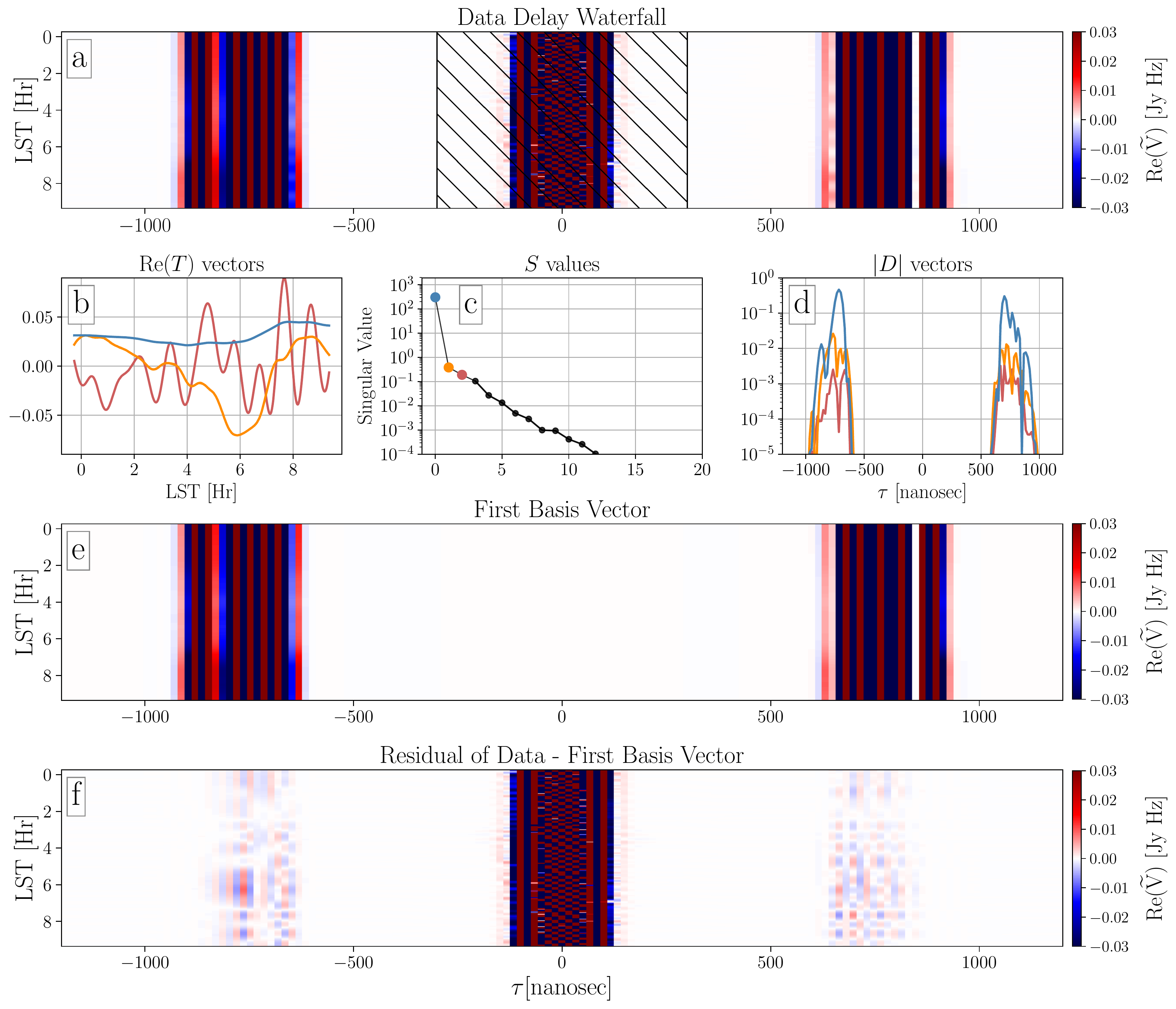}
\caption{Semi-empirical modeling and removal of cross coupling systematics from a simulated cross-correlation visibility.
{\bfseries (a)} A simulated visibility with foregrounds (center) and cross coupling systematics (left \& right).
The hatched region of $|\tau| < 300$ ns is assigned zero weight before taking the SVD.
{\bfseries (b, c, d)} The resulting $\mathbf{T}$ modes, singular values and $\mathbf{D}$ modes after factorization via SVD. The $\mathbf{D}$ modes are artificially offset for visual clarity.
{\bfseries (e)} The outer product of the first $\mathbf{T}$ and $\mathbf{D}$ mode multiplied with its singular value yields the first basis vector having the shape of the original data matrix.
{\bfseries (f)} The difference of the systematic model and the original data shows decent subtraction of the systematic, but isn't enough to completely remove it from the data.}
\end{figure*}

Our semi-empirical approach starts with the cross-correlation visibility Fourier transformed across frequency, $\widetilde{V}_{12}$, such that it is in the time and delay domain.
If we think of the visibility as a 2D rectangular matrix, we can use Singular Value Decomposition (SVD) to decompose the matrix as
\begin{align}
\label{eq:svd}
\widetilde{\mathbf{V}} = \mathbf{T} \mathbf{S} \mathbf{D}^\dagger,
\end{align}
where $\mathbf{T}$ is a unitary matrix containing basis vectors (also referred to as eigenmodes) across time, $\mathbf{D}$ is a unitary matrix containing basis vectors across delay, and $\mathbf{S}$ is a diagonal matrix containing the weight (or their singular values) of each mode in the data.
There may be components of our data matrix, $\widetilde{V}$, that are inherently low-rank, like a slowly time-variable systematic for example.
Thermal noise in the visibility, on the other hand, occupies the full-rank of the data matrix.
SVD can help us model and pull out the low-rank components of the matrix, thus providing an approach for systematic removal on a per-baseline basis.

Our SVD-based systematic removal algorithm operating on an individual visibility takes the following steps:
\begin{itemize}
\item[1.] Fourier transform the visibility waterfall to delay space.
\item[2.] Apply a rectangular band-stop window across delay to down-weight foregrounds at low delays.
\item[3.] Decompose the visibility via SVD.
\item[4.] Choose the first $N$ modes to describe the systematic and truncate the rest.
\item[5.] Take the outer product of the remaining $\mathbf{T}$ and $\mathbf{D}$ modes to form $N$ data-shaped templates.
\item[6.] Multiply each template with their corresponding singular value in $\mathbf{S}$ and sum them to generate the full time and delay-dependent systematic model.
\item[7.] Fourier transform the systematic model from delay space back to frequency space and subtract it from the data.
\end{itemize}
A demonstration of this process on simulated visibilities is shown in  \autoref{fig:sim_xtalk_sub}.
In this example, the simulation contains foregrounds and cross coupling systematics, but is free of both thermal noise and EoR components.
We start with a simulated visibility spanning roughly 9 hours in LST with 1000 time bins and spanning a bandwidth from 120 -- 180 MHz with 256 frequency bins.
By Fourier transforming it to the delay domain (a), we see that foregrounds are confined to low delays while cross coupling systematics span a wide range of delays at positive and negative delay offsets.
Applying a band-stop windowing function across delay before taking the SVD (hatched region) assigns zero weight to delay modes dominated by foreground signal at $|\tau| < 300$ ns.
The result of the SVD shows significant isolation of the information content of the systematic in the visibility into the first eigenmode (c).
The first time eigenmode (b) indeed shows it to be slowly time variable, as we would expect for a cross coupling systematic (see \autoref{sec:prelim}).
The outer product of the first time and delay eigenmodes multiplied with their singular value yields a systematic template with the shape of our original data matrix (d).
Taking this single template as our systematic model (equivalent to setting $N=1$) and subtracting it from the data yields the systematic-subtracted data (f).

\begin{figure*}
\centering
\label{fig:bl_frfilters}
\includegraphics[scale=0.55]{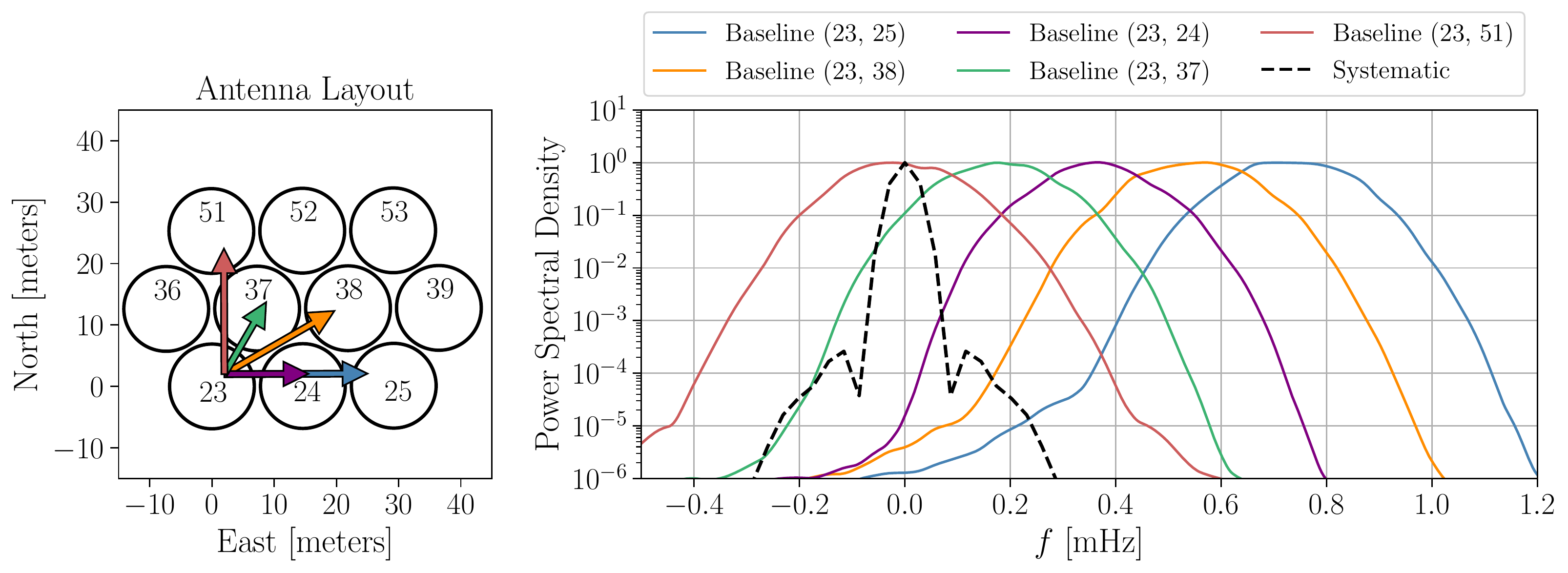}
\caption{Peak-normalized PSDs in fringe-rate space of an uncorrelated Gaussian EoR sky model for HERA baselines at 120 MHz.
{\bfseries Left:} A subset of the HERA array showing its shortest baselines arranged in a hexagonal pattern.
The arrows denote four unique baseline orientations.
{\bfseries Right:} Peak-normalized PSDs describing how EoR sky signal populates the inteferometric visibility
in the fringe-rate domain, as well as a cross-coupling systematic (black dashed).}
\end{figure*}

We see in (d) that a cross coupling model with $N=1$ provides good subtraction of the systematic, but is not enough to completely remove it: based on \autoref{fig:sim_xtalk_sub} (c) we can see that it provides effectively three orders of magnitude of suppression, which, depending on the inherent amplitude of the systematic, may or may not be enough to push it below fiducial EoR levels.
We can remove more and more of the systematic by increasing the number of SVD modes we incorporate into the systematic model.
However, as is the case with empirically-based models, this has the side-effect of possibly introducing structure from other components of the visibility that we do not want in our systematic model, such as the EoR itself.
If EoR signal was somehow soaked up by our systematic model, then by subtracting the model from the data we are inducing EoR signal loss, which is highly undesirable.

\begin{deluxetable}{ccc}[h!]
\tabletypesize{\footnotesize} 
\tablewidth{0pt} 
\tablecaption{
EoR Visibility Power Bounds
\label{tab:fr_bounds}
}
\tablehead{Baseline Length & Baseline Angle & 99\% Power Bounds}
\startdata 
29.2 meters & 0$^\circ$ & 0.46 $<f<$ 0.95 mHz \\[.1cm]
25.3 meters & 30$^\circ$ & 0.31 $<f<$ 0.77 mHz \\[.1cm]
14.6 meters & 0$^\circ$ & 0.14 $<f<$ 0.58 mHz \\[.1cm]
14.6 meters & 60$^\circ$ & -0.05 $<f<$ 0.40 mHz \\[.1cm]
25.3 meters & 90$^\circ$ & -0.27 $<f<$ 0.21 mHz \\[.1cm]
\enddata 
\tablecomments{Power bounds are defined at $\nu=120$ MHz. Baseline angle is defined in East North Up (ENU) coordinates as $\phi^\circ$ North of East (e.g. left of \autoref{fig:bl_frfilters}).}
\end{deluxetable}


To limit EoR signal loss in the process of systematic subtraction, we can filter the systematic model to reject Fourier modes that we know hold EoR power.
In general, if a signal occupies the visibility in the fringe-rate domain with variance given by $\sigma(f)^2$ and we enact a filter on it by multiplying by a weighting function $w(f)$, then the total power of the signal before filtering, $P_{\rm before}$, can be related to the total power after filtering, $P_{\rm after}$, as
\begin{align}
\label{eq:signal_atten}
\frac{P_{\rm before}}{P_{\rm after}} = \frac{\int df\ \sigma(f)^2}{\int df\ \sigma(f)^2w(f)^2}.
\end{align}
Therefore, if we know statistically how the EoR will populate the visibilities in the fringe-rate domain--in a sense deriving their power spectral density functions (PSDs)--we can construct a Fourier filter that is tailored to reject Fourier modes in the systematic model that we know hold EoR power.

\begin{figure*}
\centering
\label{fig:cross_Npc_trials}
\includegraphics[scale=0.5]{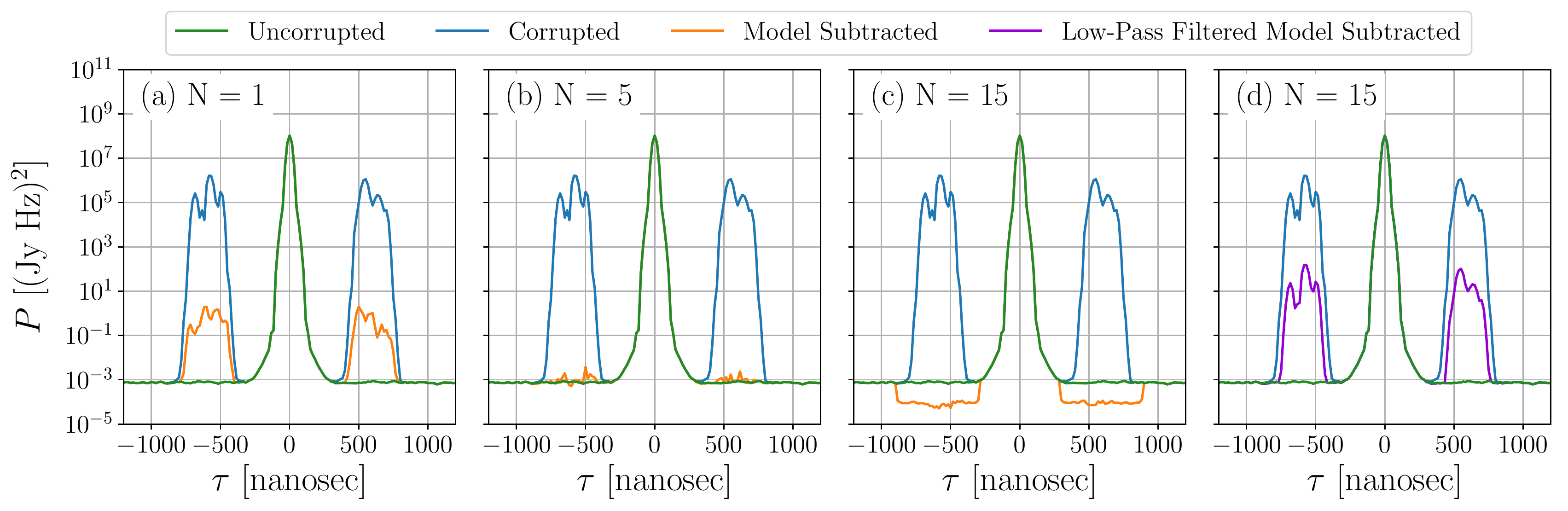}
\caption{Cross coupling systematic removal on simulated EoR + foreground visibilities for a 15-meter East-West baseline with various choices of $N$. We show the visibility amplitude averaged over LST for the uncorrupted data (green), the data corrupted by a cross-coupling systematic (blue) and the systematic-model subtracted data (orange) for $N=1, 5,\ \&\ 15$ (a, b \& c respectively).
In the last panel, we show the result of low-pass filtering the SVD $\mathbf{T}$ modes before forming the full systematic model and subtracting it from the data. For the baseline at hand, we do this with a fringe-rate cutoff of $f_{\rm max} = 0.14$ mHz (\autoref{tab:fr_bounds}).
This shows that by low-pass filtering the systematic model, we can constrain it such that it removes the systematic as much as possible while not attenuating the EoR, and is therefore optimal even if it leaves some of the systematic in the data.}
\end{figure*}

This is closely related to the optimal fringe-rate formalism outlined in \citet{Parsons2016}.
In \autoref{fig:bl_frfilters} we show peak-normalized power spectral densities (PSD) of an EoR sky model at 120 MHz for various baselines in the array, which describe the relative amount of signal power occupied by different fringe rates.
We derive these via ensemble simulations of the same EoR sky while varying the initial seed (\autoref{sec:vis_sims}).
The PSD of a simulated HERA cross coupling systematic is also plotted (black dashed).
While centered at a fringe-rate of 0 mHz, the systematic has a tail that extends out to negative and positive fringe-rates, the latter being where most of the power from the EoR also lies.
Conversely, while most of the EoR power lies at positive fringe-rates for many baselines, some of its power also extends to zero and negative fringe-rates.
Baselines that are longer along the East-West direction have more EoR power pushed to higher fringe-rates and thus are more naturally separated from cross coupling systematics, while baselines that are purely North-South in orientation see an EoR signal that is centered at $f\sim0$ mHz, and almost completely overlaps the cross coupling systematic.

Each curve in \autoref{fig:bl_frfilters} integrated out to 99\% of their total area yields the domain in fringe-rate space where 99\% of the EoR power is contained in the visibility.
We tabulate these bounds in \autoref{tab:fr_bounds} for a few HERA baselines at $\nu=120$ MHz.
If we low-pass filter any of the visibilities in \autoref{fig:bl_frfilters} in fringe-rate space by applying a symmetric top-hat filter with a maximum extent $f_{\max}$ given by these lower bounds, then \autoref{eq:signal_atten} tells us we will retain 99\% of the EoR power in the data after filtering, which for our purposes is an effectively lossless operation given other more dominant sources of error.
This result means that our ability to safely remove cross coupling systematics is baseline-dependent:
for baselines with large East-West lengths (e.g. blue), we can filter out the vast majority of the systematic without attenuating the EoR.
For shorter baselines (e.g. green), we may only be able to remove part of the systematic, and for baselines oriented along the North-South direction (e.g. red), we may not be able to remove any cross coupling systematics, if they exist.

Armed with the ability to filter an arbitrary signal without attenuating its EoR component, we can return to the problem of choosing the appropriate number of eigenmodes to use in describing a cross coupling systematic in the data.
We noted in \autoref{fig:sim_xtalk_sub} that by increasing the number of SVD eigenmodes used to describe our systematic model, we might be able to remove more of the systematic from the data.
\autoref{fig:cross_Npc_trials} proves this on a simulated visibility, now simulated with an EoR and foreground component (green) corrupted by a cross coupling systematic at high delays (blue).
We also show SVD-based systematic removal with increasingly more eigenmodes used to describe the systematic (orange).
We can see that going from $N=1$ (a) to $N=5$ (b) enables us to subtract more of the systematic from the data.
However, at some point we expect low-level eigenmodes to be influenced by other components of the data, such as EoR, foregrounds or noise, which raises the possibility of removing those components along with the systematic.
This is shown in (c), where with $N=15$ we have over subtracted the systematic and caused signal loss of EoR power at high delays.
One might conclude from this that $N=5$ is the ``sweet spot'' choice for the number of eigenmodes to use, but this choice is conditional on the relative amplitude between EoR and systematic: without knowing the amplitude of EoR in the data a priori, we have no way of knowing the appropriate number of eigenmodes to use that would enable us to subtract the systematic without attenuating EoR.
This makes the algorithm as originally described in effect unusable, because it is an operation that is dangerously lossy to EoR signal.

The solution to this problem is to apply a low-pass time filter to the systematic model that is tailored to reject fringe-rate modes occupied by the EoR.
Specifically, we can apply a filter to the systematic model $\mathbf{T}$ matrix that only keeps structure below some pre-defined maximum fringe-rate, $f_{\rm max}$, such that in the process of subtracting it from the data, all fringe-rates $|f| > |f_{\rm max}$ are left unaffected.
For example, if we could tolerate a maximum of 1\% attenuation of EoR power in the process of systematic removal, then $f_{\rm max}$ would be set at the lower bounds tabulated in \autoref{tab:fr_bounds}.
The result of applying such a filter to the SVD eigenmodes is demonstrated in \autoref{fig:cross_Npc_trials} (d), which shows the systematic-subtracted data with $N=15$ having first applied a low-pass filter to $\mathbf{T}$.
Although a significant amount of systematic remains, we can now be confident that we have not attenuated the EoR signal in the data, even while using a large number of eigenmodes to describe the systematic.

In this section we have argued and shown via sky and instrument simulations that we can construct a cross coupling model that removes the vast majority of the systematic while remaining lossless to the EoR signal (for certain baseline orientations).
In real data, however, the fidelity of this model will be fundamentally limited by the thermal noise floor of the observation, as is the case for any signal term modeled on a per-baseline basis.
If the cross coupling systematic is truly baseline-dependent and is uncorrelated between baselines, then the residual systematic term will integrate down like thermal noise when we combine visibilities and we would not expect it to re-appear in the integrated power spectra.
In HERA, for example, there is evidence that the observed cross coupling systematics are at least partially uncorrelated between baselines \citep{Kern2019b}.

\section{Signal Loss}
\label{sec:signal_loss}
We have thus far presented an overview of instrumental systematics that can hinder if not prohibit the detection of the EoR for current and future \tocm intensity mapping surveys, and have outlined algorithms for modeling and removing them from the data.
However, any experiment that wishes to use a systematic removal technique on the data must show that the subtraction did not attenuate the desired signal in the data.
In other words, one must quantify and account for possible sources of signal loss in a data reduction pipeline.
In this work, we use signal loss to refer specifically to the inadvertent subtraction of sky signal (EoR or foreground) from the visibilities.

Quantifying signal loss can be done in a variety of ways depending on the nature of the algorithm one wants to test \citep{Cheng2018, Mouri2019}.
In general, however, we can quantify the amount of signal loss induced by an algorithm by generating two identical mock datasets, introducing a systematic to one of them, attempting to remove it, and then comparing the end-result power between the two datasets.
For this analysis, we generate mock observations using the same simulations used in \autoref{sec:modeling}, but now in addition to diffuse foregrounds and systematics we introduce an EoR component.
The EoR sky model is an uncorrelated Gaussian random field across both the spatial and frequency axes with a variance of 25 mK$^2$ (see \autoref{sec:vis_sims} for details).

We simulate the EoR and foreground visibilities separately, and then assign their sum as $V_1$.
Next we create and add in a systematic visibility and assign their sum as $V_2$.
Lastly, we create a visibility model of the systematic using our algorithms presented above, remove it from the data and assign the residual as $V_3$, which can be summarized as follows:
\begin{align}
V_1 &= V_{\rm eor} + V_{\rm fg} \nonumber \\
V_2 &= V_1 + V_{\rm sys} \nonumber \\
V_3 &= F(V_2) = V_2 - V_{\rm mod}
\end{align}
where $F$ is a systematic removal algorithm whose signal loss properties we would like to quantify, and whose effect is to subtract a model of the systematic, $V_{\rm mod}$, from the corrupted visibility.
Note we do not include a thermal noise term, which is done so that we can probe the signal loss properties of the algorithms down to the extremely weak levels of a fiducial EoR signal.

Each visibility has an associated total power, which is a real-valued quantity and can be calculated as the square of the Fourier transformed visibility.
In the ideal EoR + foreground case, this is
\begin{align}
\label{eq:P1}
P_{\rm 1} &= \widetilde{V}_{\rm 1}\widetilde{V}_{\rm 1}^\ast \nonumber \\
&= P_{\rm eor} + P_{\rm fg} + 2\Re(P_{\rm eor,fg})
\end{align}
where $\widetilde{V}$ signifies the visibility Fourier transformed from frequency to delay space, and $P_{\rm eor,fg}$ represents the cross-power between $\widetilde{V}_{\rm eor}$ and $\widetilde{V}_{\rm fg}$.
Similarly, we can write the power of the systematic-subtracted visibility as,
\begin{align}
\label{eq:P3}
P_{\rm 3} =&\ \widetilde{V}_3\widetilde{V}_3^\ast \nonumber \\
=&\ P_{\rm eor} + P_{\rm fg} + 2\Re(P_{\rm eor,fg}) \nonumber \\
&+ P_{\rm sys} + P_{\rm mod} - 2\Re(P_{\rm sys,mod}) \nonumber \\
&+ 2\Re(P_{\rm eor,sys}) - 2\Re(P_{\rm eor,mod}) \nonumber \\
&+ 2\Re(P_{\rm fg,sys}) - 2\Re(P_{\rm fg,mod}).
\end{align}
In the case where we have \emph{perfectly subtracted} the systematic from the visibility (i.e. $V_{\rm mod} = V_{\rm sys}$), we see that the systematic power terms cancel with the model power terms such that, not surprisingly, we get that $P_{3} = P_{1}$.
In the case where we have \emph{imperfectly subtracted} the systematic--either by incorrect estimation of its phase and/or amplitude--the cross terms no longer cancel.
What this means for the total power of the resultant visibility, $P_{3}$, depends on how well-matched the model visibility is to the systematic, in addition to the relative inherent amplitude of the sky signal versus the systematic.
For example, in the case of an imperfect systematic model, then we can see that this will always be true: $P_{\rm sys} + P_{\rm mod} > 2\Re(P_{\rm sys,mod})$, meaning their difference results in excess power.
Whether or not the EoR and foreground cross term residuals in \autoref{eq:P3} result in overall positive or negative power depends on how well matched the systematic model is to either EoR or foregrounds.

%
\begin{figure*}
\centering
\label{fig:ref_sig_loss}
\includegraphics[scale=0.48]{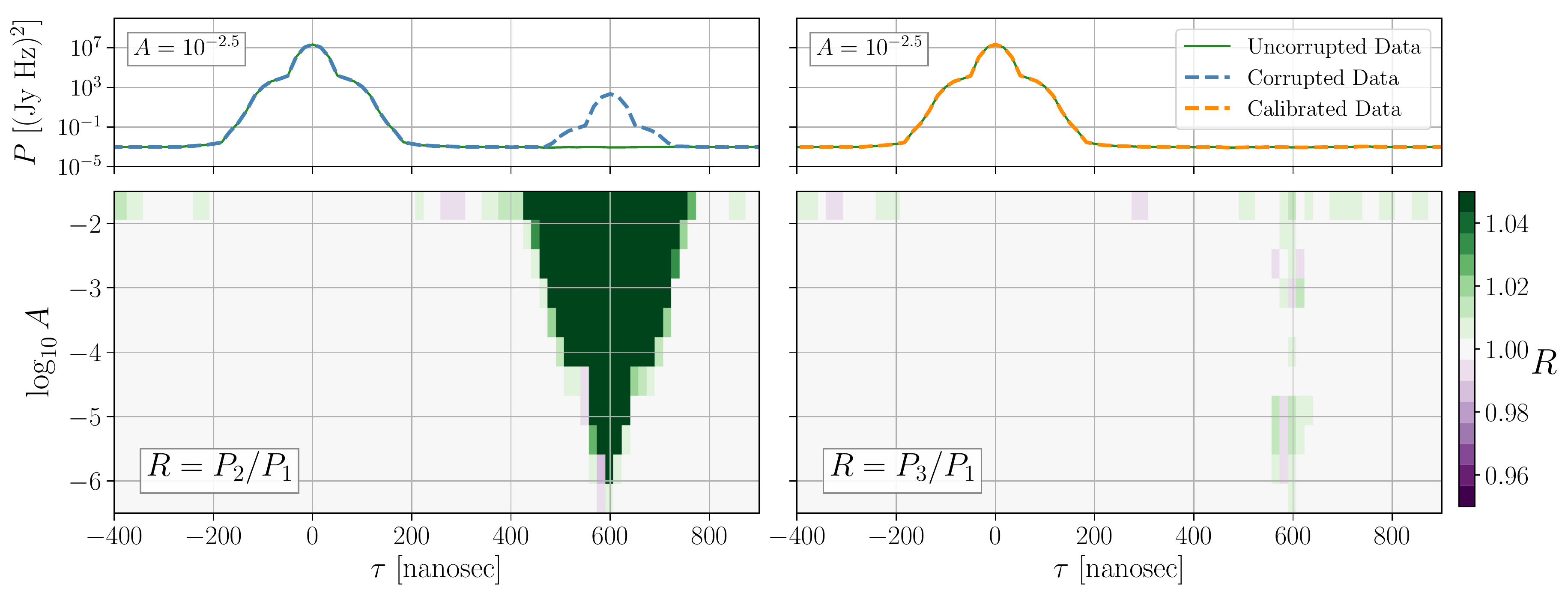}
\caption{Signal loss trials of reflection calibration with noise-free, foreground + EoR simulated visibilities.
{\bfseries Top Row:} Power spectra of the corrupted visibility $V_2$ (blue-dashed), the uncorrupted visibility $V_1$ (green-solid) and the calibrated visibility $V_3$ (orange-dashed) from a signal loss trial with a reflection amplitude of $A = 10^{-2.5}$.
{\bfseries Bottom Row:} Heatmaps of the signal loss $R$ metric computed for the corrupted data (left) and the calibrated data (right) as a function of reflection amplitude (y-axis) and delay (x-axis). The residual fluctuation about $R=1$ in the right panel is encompassed within the $1/\sqrt{N}$ sample variance of our finite ensemble average.
}
\end{figure*}

Given this, we can construct a simple metric,
\begin{align}
\label{eq:R_metric}
R_3(\tau) = \frac{\langle P_3(\tau)\rangle}{\langle P_1(\tau)\rangle}
\end{align}
to determine whether or not a step in our data analysis induces signal loss.
Here the $\langle\rangle$ denotes an ensemble average over many realizations of the visibilities with the same kind of sky signal and systematic.
Taking the ensemble average before taking the ratio is done to ensure the power spectra are properly normalized.
Signal loss occurs anytime $R_3(\tau) < 1$, meaning that our model-subtracted visibility, $V_{3}$, has less power in it than our uncorrupted visibility, $V_{1}$.
Specifically, \emph{EoR signal loss} occurs anytime $R_3(\tau) < 1$ at delays we know to be dominated by EoR over foregrounds, such as all delays significantly outside the geometric delay of the baseline.
In the case of $R_3(\tau) > 1$, the resultant visibility is systematic limited but, importantly, is not under reporting the power in the data relative to the pure sky signal visibility.
We can also form the metric $R_2(\tau)$ using $P_2$ instead of $P_3$, which informs us of the relative amplitude of the raw systematic (without any removal) compared to the underlying sky signals.

Whether or not an algorithm is lossy in practice can depend on the relative amplitude between the signal and systematic present in the data.
As such, we need to compute the $R$ metric while varying the relative amplitude between the EoR and systematics.
\citet{Cheng2018}, for example, do this by repeatedly injecting mock EoR signals into their analysis pipeline with increasing amplitude.
In this study we take the opposite approach.
We adopt a fixed EoR amplitude consistent with rough theoretical expectations and insert systematics at amplitudes below, equal to and above the adopted EoR amplitude and compute $R$.
This approach is more consistent with what we expect to find in the real data: at certain times, frequencies, or baselines, we may find systematics to be heavily dominant, while at other times, frequencies or baselines, there may be no systematics at all.

\subsection{Signal Loss in Reflection Calibration}
\label{sec:ref_cal_sig_loss}
To test signal loss in the context of reflection calibration, we precompute the visibilities for a single diffuse foreground model and 100 independent EoR models, with each simulation spanning 8 hours of LST and a thousand individual time integrations.
Because HERA has a beam crossing time of about 1 hour, this yields an effective number of independent foreground + EoR simulations of $\sim800$.
The adopted EoR model is an uncorrelated Gaussian field across angular position and frequency with a variance of 25 mK$^2$ (\autoref{sec:vis_sims}).

A single signal loss trial takes the following steps.
First we choose a random EoR model from our library of pre-computed visibilities and add it to our foreground visibility ($V_1$).
We then make a copy of it and insert a reflection with a delay of 600 nanoseconds, a random phase, and a single amplitude across frequency using \autoref{eq:cross_corr_sig_ref} ($V_2$).
We then model the reflection in the simulated auto-correlation knowing only the approximate delay range at which it appears, and then apply the derived gain solution to the cross-correlation visibilities ($V_3$).
Next we compute power spectra of each data product ($P_1,\ P_2,\ \&\ P_3$).
We then repeat this on the order of 100 times, each with a different random EoR model, and then take their average to approximate the ensemble average in \autoref{eq:R_metric}.
We then form the $R_2$ and $R_3$ metrics as a function of time and delay--i.e. $R_3(t, \tau)$--and average over time to collapse them onto a single axis across delay.
This entire procedure produces one signal loss trial, which is defined uniquely by the amplitude, $A$, of the reflection inserted into the visibility.

\autoref{fig:ref_sig_loss} shows multiple trials for different reflection amplitudes in the range of $10^{-6}$ to $10^{-1}$.
The top row shows power spectra of each of the three visibility products as a function of delay for one trial when $A = 10^{-2.5}$.
Recall that the reflection amplitude is defined with respect to the visibility, meaning that the observed reflection amplitude in the power spectrum is $A^{2}$.
The bottom row shows a heatmap of the signal loss $R$ metric as a function of delay (x-axis) and each trial's reflection amplitude (y-axis).
The left panels shows $R_2$ and the right panel shows $R_3$, highlighting the amount of delay space that is brought down to $R\sim1$ after reflection calibration, with negligible amounts of signal loss (purple shaded regions).
Furthermore, the weak levels of fluctuating residual systematic and signal loss observed at the $\sim2\%$ level are within the $1/\sqrt{N}$ sample variance of our finite ensemble average.
For context, HERA cable reflection amplitudes are seen at around $10^{-3}$ \citep{Kern2019b}.

\subsubsection{Multi-Reflection Regime}
Above we probed for signal loss when performing reflection calibration on a single reflection that was isolated in delay space.
Next, we relax this assumption and test how the the algorithm performance and signal loss properties change when we add in more reflections,
which is relevant for any instrument with multiple cables, or with cables that have sub-reflections along the length of the cable \citep{Ewall-Wice2016b, Kern2019b}.
We choose to model the relative amplitude of these reflections as an inverse power law as a function of delay with a nominal reflection amplitude of $A = 3\times10^{3}$.
Our algorithm models and calibrates out each reflection one-at-a-time, starting with the reflection with the largest amplitude.
We do not feed the algorithm the position of each reflection (it searches for it automatically within a specified range of delays), but we do assume we know the number of reflection inherent in the data, which controls how many times we iterate the algorithm.

\begin{figure*}
\centering
\label{fig:autocorr_multiref1}
\includegraphics[scale=0.55]{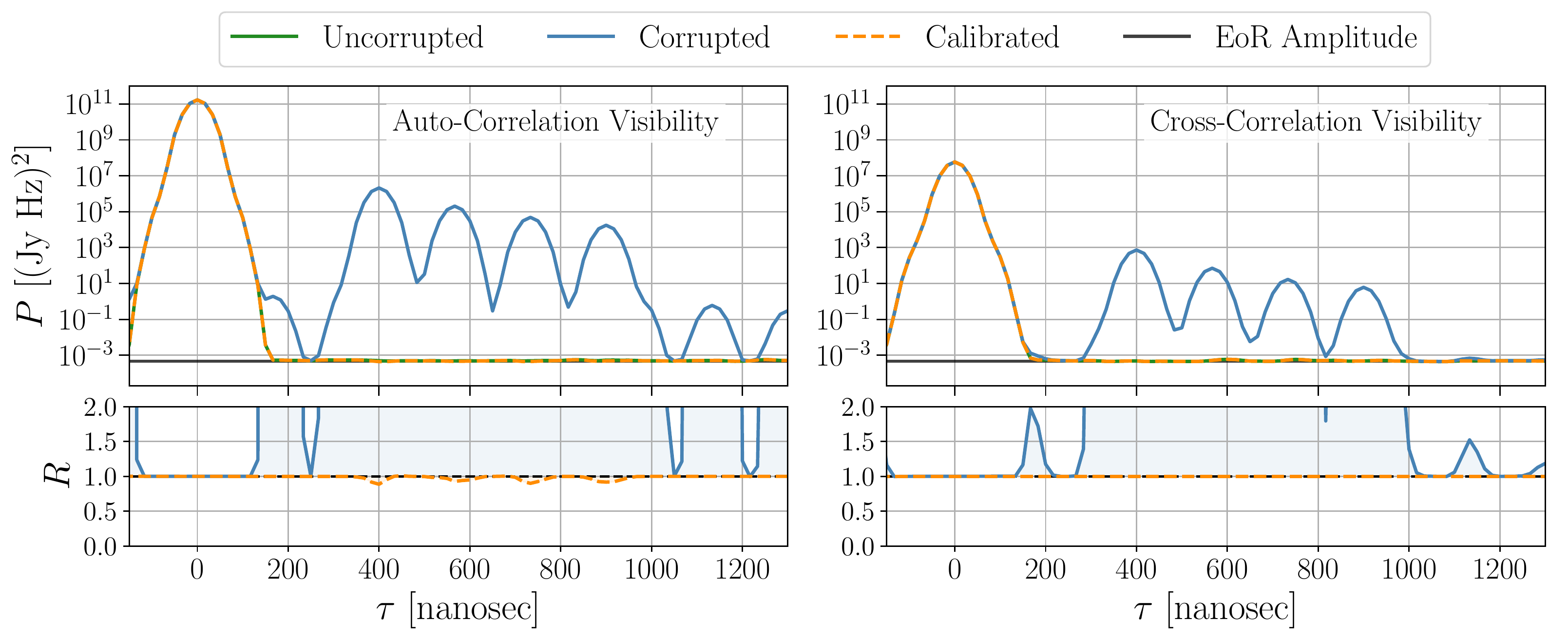}
\caption{Reflection calibration run on an auto-correlation (left) in a low-confusion, multi-reflection regime, where we then apply the resultant gains to a cross-correlation visibility with a 29-meter baseline length (right) and repeat a few dozen times. In the top panels, the (ensemble average) power spectrum of the uncorrupted data ($P_1$; green), corrupted data ($P_2$; blue) and calibrated data ($P_3$; orange-dashed) are plotted along with a line denoting the underlying EoR amplitude in the data (grey). Signal loss metrics $R_2$ (blue) and $R_3$ (orange) are shown in the bottom panels. We see that while reflection calibration can lead to a slight amount of signal loss at the reflection delays of the auto-correlation visibility (bottom-left, dashed), signal loss is not observed to an appreciable degree in the cross-correlation visibility (bottom-right, dashed).}
\end{figure*}

\begin{figure*}
\centering
\label{fig:autocorr_multiref2}
\includegraphics[scale=0.55]{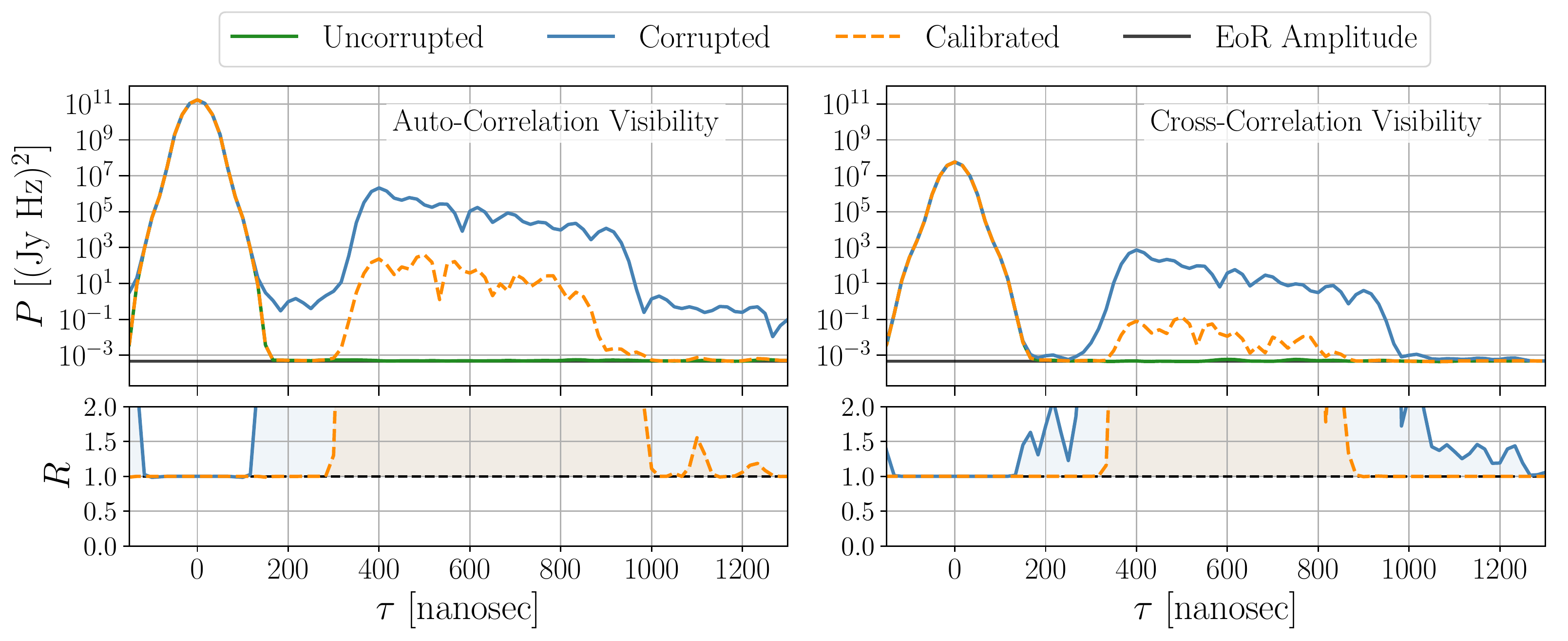}
\caption{Same figure as \autoref{fig:autocorr_multiref1} but now in a higher-confusion, multi-reflection regime. Importantly, even when reflection calibration encounters confusion in its peak finding algorithm and fails to perfectly model the reflection, it still does not induce appreciable signal loss in the cross-correlation visibility (bottom-right, orange).}
\end{figure*}

\begin{figure*}
\centering
\label{fig:cross_sig_loss_1unit}
\includegraphics[scale=0.48]{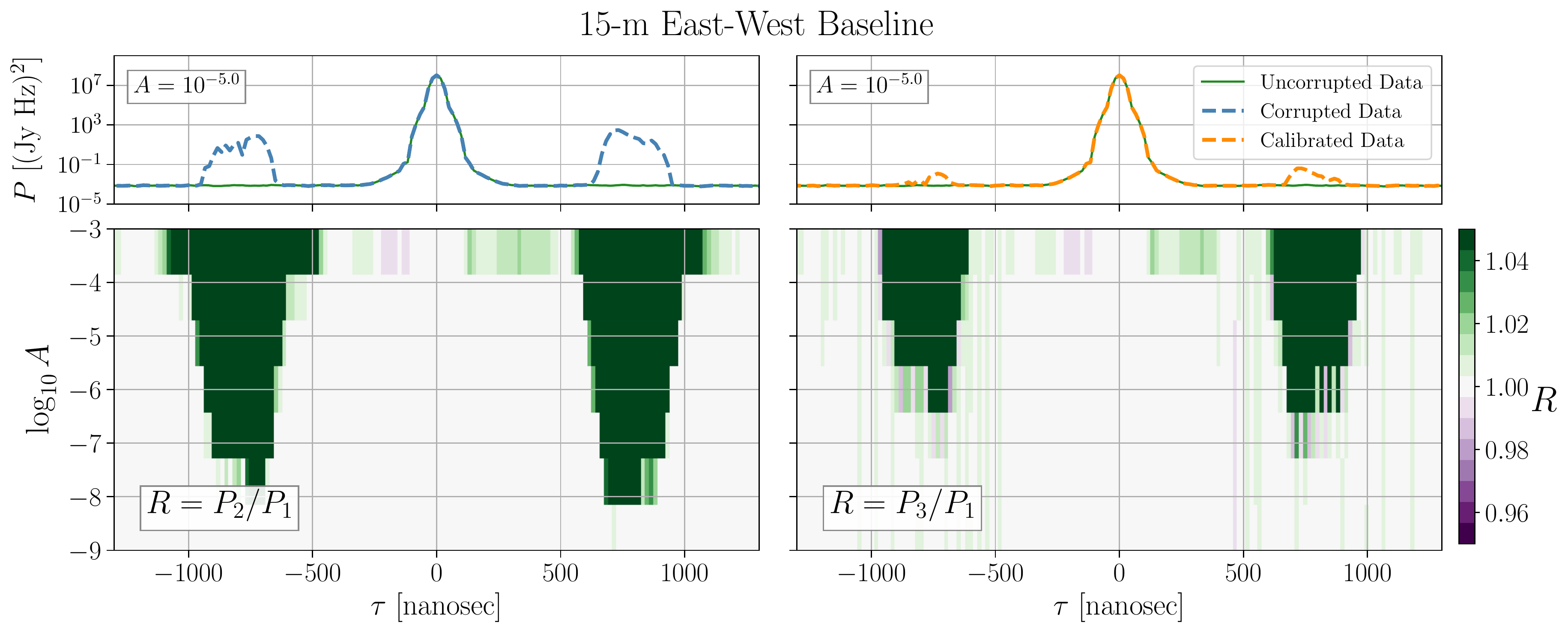}
\caption{Signal loss trials of cross coupling removal with noise-free, foreground + EoR simulated visibilities for a 15 meter East-West HERA baseline.
The systematic model is formed using 20 SVD modes and applies a low-pass time filter with an $f_{\rm max} = 0.14$ mHz.
{\bfseries Top Row:} Power spectra of the corrupted visibility $P_2$ (blue-dashed), the uncorrupted visibility $P_1$ (green-solid) and the systematic model-subtracted visibility $P_3$ (orange--dashed) from a signal loss trial with a coupling amplitude $A=10^{-4}$.
{\bfseries Bottom Row:} Signal loss $R$ metric computed for the corrupted visibility (left) and the model-subtracted visibility (right)
as a function of coupling amplitude (y-axis) and delay (x-axis), with the model-subtracted visibility (right).
No appreciable amounts of signal loss is observed, and the model-subtracted data show roughly four orders of systematic suppression in the power spectrum.
}
\end{figure*}

\begin{figure*}
\centering
\label{fig:cross_sig_loss_2unit}
\includegraphics[scale=0.48]{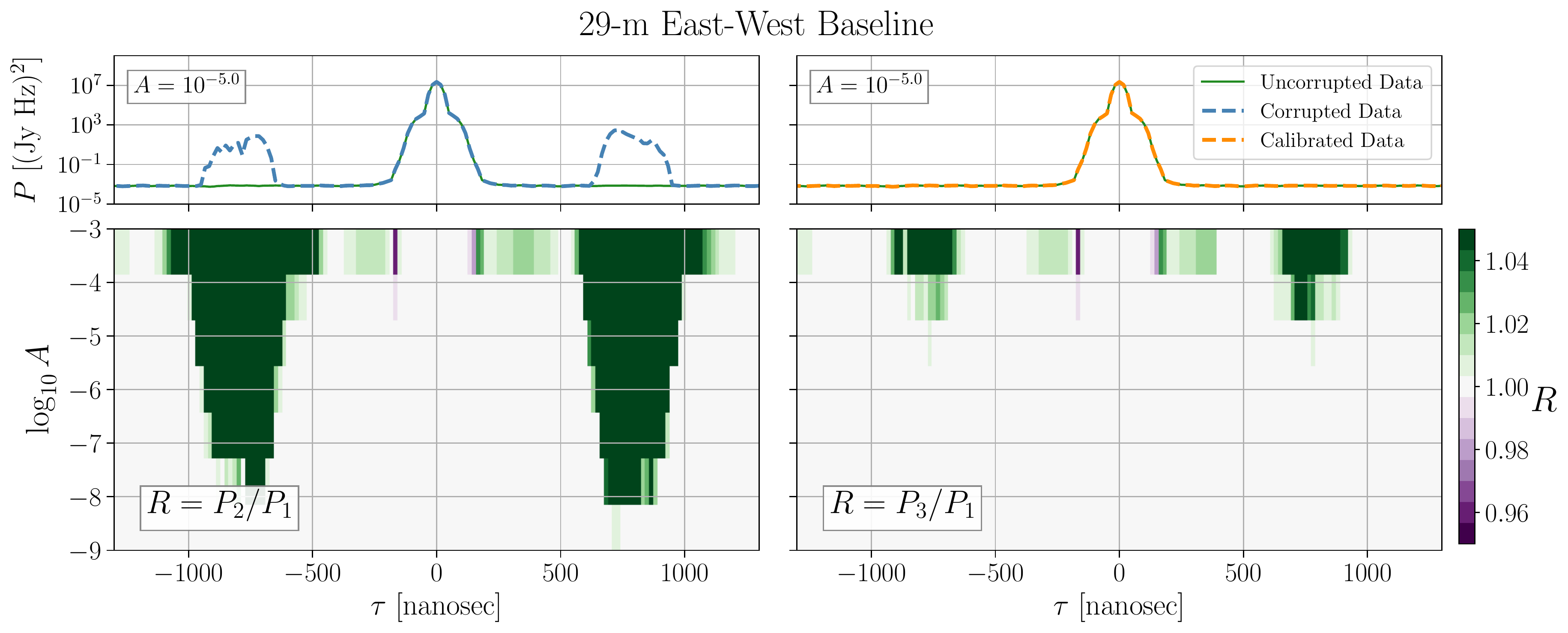}
\caption{The same signal loss trials for removal of a cross-coupling systematic as described in \autoref{fig:cross_sig_loss_1unit}, but for a 29-meter East-West baseline using a low-pass time filter with $f_{\rm max} = 0.46$ mHz. In this case we get upwards of six orders of magnitude in systematic suppression in the power spectrum.}
\end{figure*}

Our first test shown in \autoref{fig:autocorr_multiref1} involves only five reflections inserted across a relatively wide region in delay, such that they can be considered non-overlapping (or un-confused).
In the top panels, we show power spectra of the uncorrupted data ($P_3$; green), the corrupted data ($P_2$; blue) and the reflection calibrated data ($P_3$; dashed-orange), along with a line marking the EoR amplitude in the data (grey).
We show this for the auto-correlation visibility (left) and a 29-meter cross-correlation visibility (right).
In the bottom panels we show the signal loss metrics $R_2$ (blue) and $R_3$ (dashed-orange).
Recall that reflection calibration builds up a set of gains strictly from the auto-correlation, and then applies those gains to the cross-correlations.
We find that our algorithm performs exceptionally well in this regime: removing reflections down to the inherent sidelobe floor of the auto-correlation, which is more than enough to bring the systematics to the EoR level in the cross-correlation visibility.
We find that while reflection calibration can lead to slight signal loss in the auto-correlation visibility, we find no appreciable levels of signal loss in the cross-correlation.

Our next test shown in \autoref{fig:autocorr_multiref2} increases the number of reflection modes inserted into the same region, such that they become almost entirely overlapping.
In this case we can see that our algorithm fails to perfectly calibrate out the reflections in both the auto-correlation (left) and cross-correlation (right) due to the partial confusion.
Nonetheless, we still find that while imperfect reflection calibration can lead to slight signal loss in the auto-correlation, the cross-correlation is resistant to signal loss.
Reflection calibration's resistance to signal loss in the cross-correlations is perhaps not surprising, given that reflection calibration operates in an antenna-based space while EoR and other sky signals live in a baseline-based space.
Furthermore, our algorithm solely uses the auto-correlations to estimate the reflection parameters.

In total, our findings suggest that 1) our reflection calibration algorithm performs moderately well even in the many-reflection regime, and that 2) even if a broad delay region is contaminated by reflections, we can still in principle use this region for EoR measurements (or upper limits in the case of imperfect systematic removal) after reflection calibration because it does not suffer appreciable levels of signal loss.

\subsection{Signal Loss in Cross Coupling Subtraction}
In this section, we quantify signal loss for cross coupling subtraction in a similar manner.
Based on our conclusions from \autoref{sec:modeling_cross_coupling, we expect our cross coupling subtraction to be effectively lossless to EoR by construction, but testing this against ensemble signal loss trials is a good double-check and validation of our arguments.}
The rough functional form of the simulated cross coupling inserted into the visibilities is informed by cross coupling systematics observed in the HERA Phase I system \citep{Kern2019b}.
The simulations used are fundamentally the same as those in \autoref{sec:ref_cal_sig_loss}, except simulated with cross coupling systematics rather than cable reflections.
To simulate cross coupling in a cross-correlation visibility, we use \autoref{eq:coupling_cross_corr} to insert $\sim25$ modes spanning $700 < |\tau| < 900$ nanoseconds with a decaying power law as a function of $|\tau|$ for their relative amplitudes, normalized such that the maximum amplitude relative to the peak \emph{auto-correlation} foreground power equals a predefined amplitude, $A$.
In the process of systematic removal, we model the systematic with $N=20$ SVD eigenmodes and apply a low-pass fringe-rate filter on the SVD $\mathbf{T}$ modes using a Gaussian process smoothing (see \autoref{sec:fr_filtering}) with a maximum fringe-rate given by the lower bound in \autoref{tab:fr_bounds}.
We then Fourier transform the data from frequency to delay space using a 7-term Blackman-Harris window, and average across ensemble trials.
After forming the $R_2$ and $R_3$ metrics as a function of time and delay, we truncate 5\% of the time bins on either edge of the time axis before taking their time average to limit the influence of boundary effects in the smoothing process.

The result is shown in \autoref{fig:cross_sig_loss_1unit} and \autoref{fig:cross_sig_loss_2unit}, which shows signal loss trials run on a 15-meter East-West baseline and a 29-meter East-West baseline, respectively.
As expected, we see better systematic suppression for the longer baseline, where the EoR signal is inherently more isolated from the systematic in fringe-rate space.
We also see that in the case of strong coupling amplitudes we cannot completely suppress the systematic down to EoR levels; however, we can nonetheless suppress the systematic by roughly four orders of magnitude in power for the 15-meter baseline and eight orders of magnitude in power for the 29-meter baseline.
Importantly, we show that the algorithm presented does not significantly attenuate EoR in the data, with residual fluctuations about $R=1$ at the 1\% level in power.
While this result is not surprising given the fact that in \autoref{sec:modeling} we constructed our systematic model to explicitly be lossless to EoR, it is still a useful cross check on our algorithm and its implementation on the data.

\section{Summary}
In this work, we present an overview of \tocm radio survey systematics related to internal instrument coupling, including signal chain reflections and antenna cross couplings.
Such systematics will hinder cosmological surveys aiming to detect and characterize the \tocm signal from the EoR and Cosmic Dawn.
We study the temporal and spectral behavior of these systematics in simulations, and propose techniques for modeling and removing them from the data without attenuating the desired cosmological signal.
We further test the signal loss properties of these techniques with simulated HERA sky and systematic simulations.

For simulated cable reflection systematics in the absence of noise, our method can estimate its parameters to 1 part in $10^6$ and achieve suppression of over eight orders of magnitude in the power spectrum for both isolated and semi-isolated reflections.
In practice, non-ideal effects like frequency evolution in the reflection parameters will limit the performance of this technique on real data.
Through signal loss trials we show that reflection calibration, when modeled from the auto-correlation visibility, is resistant to EoR signal loss in the cross-correlation visibilities.

Antenna cross couplings are another category of systematics that we address in this work.
We present an SVD-based modeling technique that, when low-pass filtered along the time axis, can suppress cross coupling systematics by four orders of magnitude in the power spectrum for HERA's short East-West baseline, and eight orders of magnitude or more for baselines of longer projected East-West separation in the absence of noise.
We show that we can tailor our systematic model to minimize its susceptibility to inducing EoR signal loss, and additionally prove this with signal loss trials. This result has critical implications for enabling \tocm experiments to mitigate systematics that would otherwise hinder if not prohibit them from making a robust detection of the EoR signal.
A companion paper \citep{Kern2019b} applies these methods to HERA Phase I data as a method demonstration, showing we can suppress reflection and cross coupling systematics down to the array-integrated noise floor of the data for a single nightly observation.

\

The authors would like to thank Daniel Jacobs, Aaron Ewall-Wice, Miguel Morales, Jonathan Pober, Bryna Hazelton, James Aguirre and Adrian Liu for helpful discussions related to this work. This material is based upon work supported by the National Science Foundation under Grant Nos. 1636646 and 1836019 and institutional support from the HERA collaboration partners. This research is funded in part by the Gordon and Betty Moore Foundation. J. S. D. acknowledges the NSF Astronomy and Astrophysics Postdoctoral Fellowship (NSF Award \#1701536). A. E. L. would like to acknowledge NASA Grant 80NSSC18K0389.

\appendix{}

\section{Simulated Visibilities with \texttt{healvis} and \texttt{hera\_sim}}
\label{sec:vis_sims}
We use the numerical visibility simulation package \texttt{healvis}\footnote{\url{https://github.com/RadioAstronomySoftwareGroup/healvis}} to compute mock observations of foreground and EoR sky models.
\texttt{healvis} numerically integrates the measurement equation (\autoref{eq:ME}) by representing the sky and direction-dependent antenna primary beam response as HEALpix maps \citep{Gorski2005}, and summing their product with the baseline fringe pattern to compute the visibility.
The mechanics of \texttt{healvis} as a simulator is described in more detail in \citet{Lanman2019}.
Our simulations use HEALpix sky maps with NSIDE = 128 and use a frequency and angular-dependent electromagnetic simulation of the HERA dish and feed primary beam response \citep{Fagnoni2019}.
The adopted beam model does not include mutual coupling effects.
The beam has been smoothed across frequency to limit excess spectral structure above 250 ns, mimicking an idealized HERA beam response.
We simulate visibilities for HERA baselines of various orientations and separations ranging from 0 meters in separation (i.e. the auto-correlation) out to 60 meters in separation.
The sky resolution provided by a HEALpix NSIDE = 128 map is roughly ten times smaller than the fringe wavelength of the longest baseline in consideration at the highest simulated frequency of $\nu=180$ MHz.

\begin{figure*}[h]
\centering
\label{fig:healvis_skies}
\includegraphics[scale=0.48]{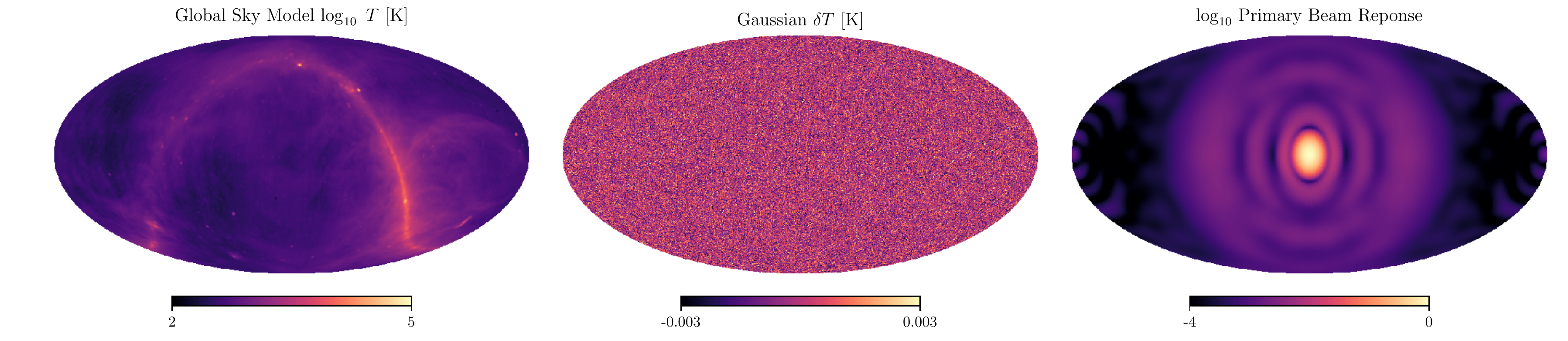}
\caption{HEALpix sky maps at $\nu=120$ MHz used for simulating diffuse foregrounds (left) and an uncorrelated EoR field (center).
The antenna primary beam response (right) is taken from an electromagnetic simulation of the HERA dish and feed \citep{Fagnoni2019}.}
\end{figure*}

\begin{figure*}[h]
\centering
\label{fig:healvis_vis}
\includegraphics[scale=0.5]{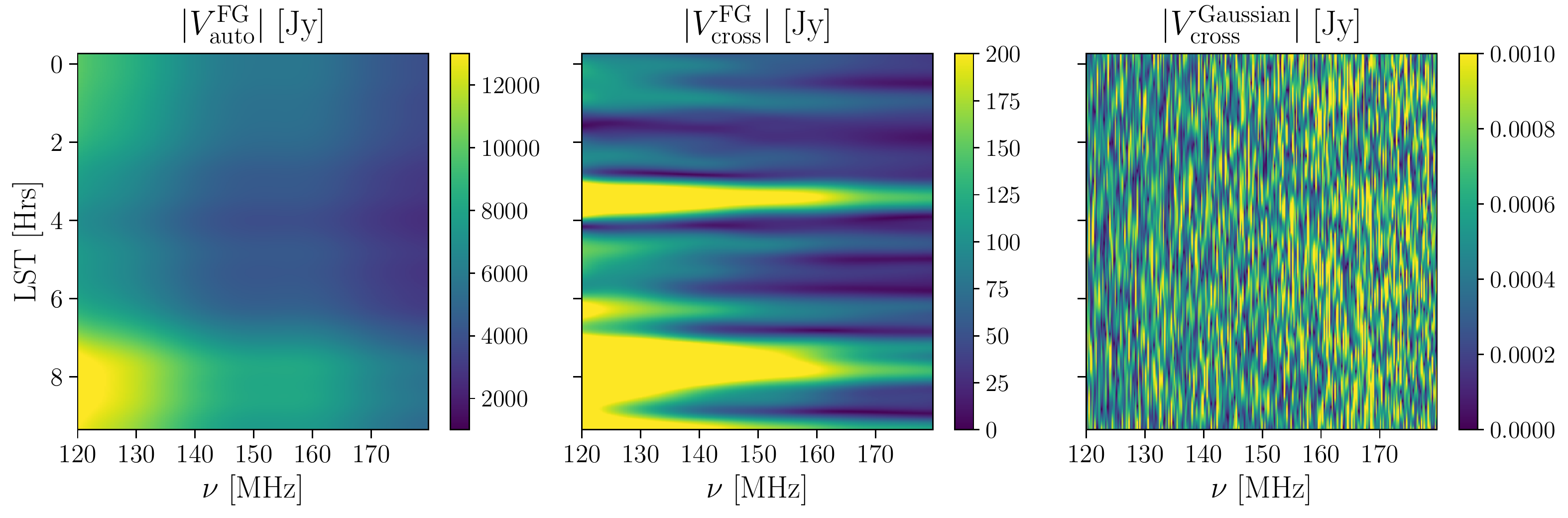}
\caption{Mock visibilities computed with \texttt{healvis} and \texttt{hera\_sim} showing (left) an auto-correlation visibility of diffuse foregrounds, (center) a cross-correlation visibility of diffuse foregrounds, and (right) a cross-correlation visibility of an EoR model.}
\end{figure*}

Foreground visibilities are simulated with HEALpix maps of the diffuse, low-frequency radio sky from the  \texttt{PyGSM} package\footnote{\url{https://github.com/telegraphic/PyGSM}}, which is a repackaging of the original 2008 Global Sky Model \citep[GSM;][]{Oliveira2008}.
We simulate a bandwidth spanning 120 -- 180 MHz with 256 channels and an LST range of roughly 0 - 8 hours with a time cadence of 30 seconds (about 1000 time bins), which corresponds to the transit of the cold part of the radio sky.
\autoref{fig:healvis_skies} shows the diffuse radio sky from the GSM, a mock EoR realization, and the adopted antenna primary beam response at $\nu=120$ MHz.
The EoR model is constructed as an uncorrelated $\delta T$ field with a variance of 25 mK$^2$, consistent with fiducial expectations for the signal at EoR redshifts \citep{Mesinger2011}.
Its Fourier correlations are modeled as a flat spectrum in $P(k)$: while fiducial EoR models tend to favor EoR as a roughly flat spectrum field in $\tfrac{k^3}{2\pi^2}P(k)$, for the purpose of using these simulations as mock visibilities for validation and testing we require only a semi-realistic EoR model, and believe this difference to have negligible impact on our results.
The pixel size of an NSIDE 128 map corresponds to a transverse comoving length scale of $\sim70$ cMpc at $z=8$, which is larger than the size scales where EoR is correlated during the beginning and middle of reionization.
The frequency axis is simulated with a channel resolution of 234 kHz, which at redshift $z=8$ corresponds to a comoving length scale of $\sim4$ cMpc, which is roughly the scale at which our uncorrelated Gaussian field approximation begins to break down.
However, we also believe this to have a negligible impact on our performance and signal loss tests: the most sensitive part of our analysis is the computation of the EoR PSD functions (\autoref{fig:bl_frfilters}) which relies on the time correlations of the EoR model, not its frequency correlations.
\autoref{fig:healvis_vis} shows simulated \texttt{healvis} visibilities of the diffuse foreground (left) and EoR model (center) described above.

We also use the visibility simulation toolbox \texttt{hera\_sim}\footnote{\url{https://github.com/HERA-Team/hera_sim}} to model signal chain reflection and cross coupling systematics.
\texttt{hera\_sim} is a general purpose toolbox for creating mock observations with realistic instrumental and environmental effects, like thermal noise, reflections and RFI. 
For inserting systematics into the data, \texttt{hera\_sim} uses the equations outlined in \autoref{sec:prelim}, in particular \autoref{eq:auto_corr_sig_ref} and \autoref{eq:coupling_cross_corr}.
For the signal loss trials described in \autoref{sec:signal_loss}, we simulate 100 independent EoR visibilities coming from EoR sky maps generated with unique random seeds, and add the GSM foreground visibility to each one.
We then add in the relevant systematic to be tested, and use this set of EoR + foreground + systematic visibilities to perform the ensemble average needed for quantifying signal loss.

\begin{figure*}
\centering
\label{fig:healvis_frate_fit}
\includegraphics[scale=0.55]{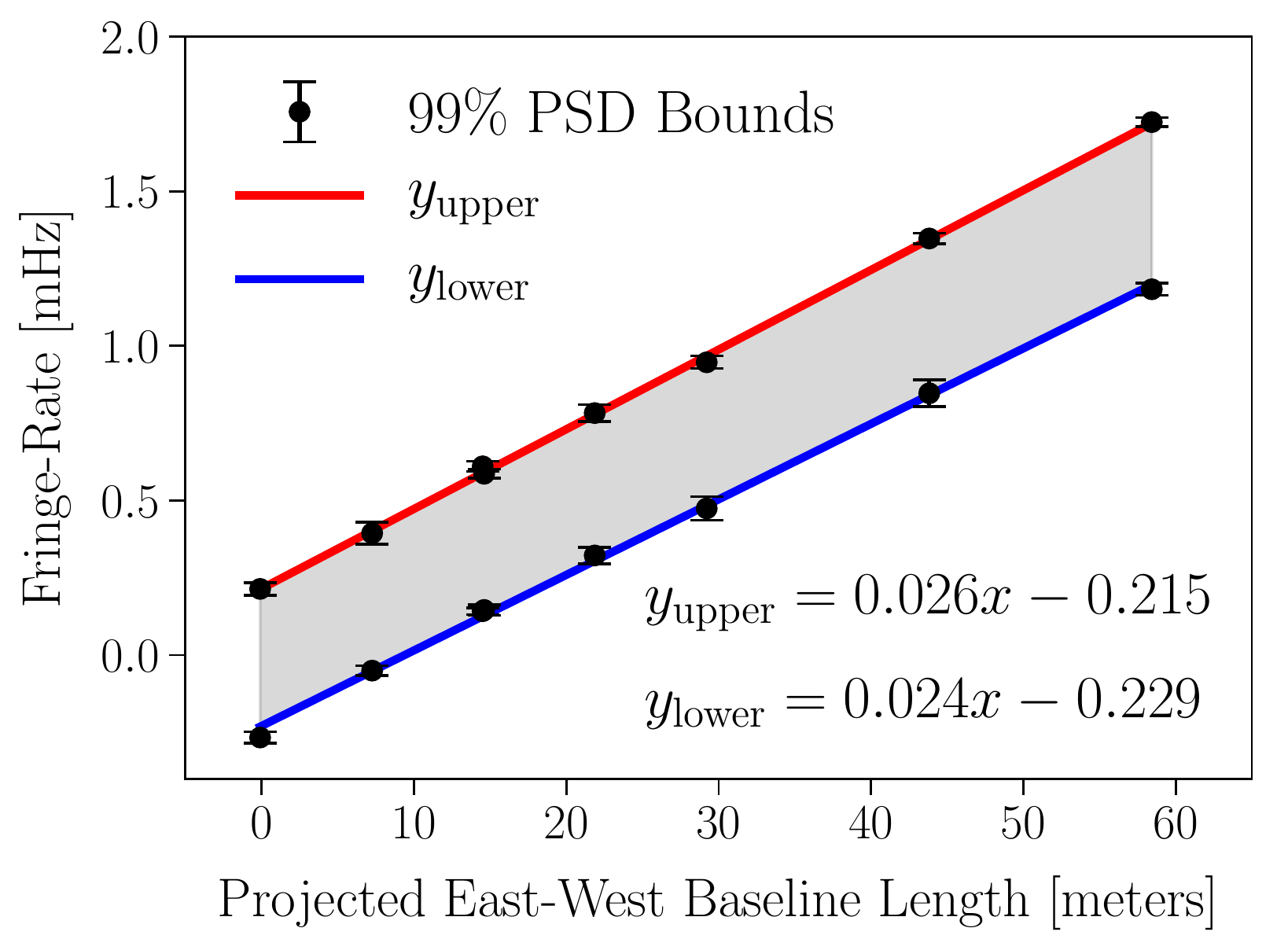}
\caption{Power spectral density (PSD) bounds of EoR sky models for HERA baselines in fringe-rate space at $\nu=120$ MHz. The PSD curves are shown in \autoref{fig:bl_frfilters}, and the bounds quoted correspond to 99\% of the total power. Best-fit lines are shown for extrapolation to other HERA baselines.}
\end{figure*}

\subsection{EoR Power Spectral Density Functions for HERA}
We use \texttt{healvis}-simulated HERA observations to calculate the expected power spectral density (PSD) functions of an EoR-like signal for HERA visibilities.
At any given time, a point source locked to the celestial sphere generates a complex sinusoid in the inteferometric visibility with a time period determined by its position on the sky and the fringe profile of the baseline at hand.
Over a short time interval, its Fourier transform across time is nearly a delta function at a fringe-rate set by the inverse of its time period.
Thus, over short time intervals, points on the sky map to specific fringe-rates in the visibility \citep{Parsons2016}, which is also related to the $m$-mode analysis \citep{Shaw2014}.
The analytically-derived ``optimal fringe-rate filter'' in \citet{Parsons2016} is a filter that minimizes the noise-component of the visibility-based power spectrum errorbars, and is related to the PSD of EoR-like signals in the visibility.
However, the analytic derivation hinges crucially on the assumption that the beam crossing time is much longer than the fringe-crossing time for a point source on the sky, which was a fairly valid assumption for a wide-field experiment like PAPER.
This is not the case for HERA, which has both shorter baselines with wider fringes and also a more compact primary beam.
Therefore, calculating the power spectral density of an EoR-like sky signal is more easily done numerically.
We do this by generating over 100 independent EoR sky models with the same variance but different initial seeds.
We perform HERA visibility simulations of each sky using \texttt{healvis}, Fourier transform them from LST into the fringe-rate domain and then take their absolute value and average across each realization.
The square of these profiles is shown in \autoref{fig:bl_frfilters}, which represents a numerically-derived PSD of a theoretical EoR-like signal in HERA visibilities.
These profiles allow us to tailor visibility Fourier filters to do things like minimize EoR signal loss in systematic subtraction, or to minimize the thermal noise component in the power spectrum errorbars.
We tabulate the 99\% power bounds of these curves in \autoref{tab:fr_bounds} for the few baselines presented in this study.
We also plot these bounds and provide a fitting formula as a function of projected East-West baseline length in \autoref{fig:healvis_frate_fit}, such that one can extrapolate these data points to other baselines in the HERA array.
Errorbars on the 99\% bounds in \autoref{fig:healvis_frate_fit} are calculated via bootstrap resampling over the independent realizations \citep{Efron1994}.

\section{Comparison of Cross Coupling Removal Techniques}
\label{sec:fr_filtering}
In this section, we compare techniques for suppressing slowly-fringing systematics (e.g. crosstalk), including the one presented in this work in \autoref{sec:modeling}.
Specifically, we compare the technique from the PAPER collaboration of convolving a finite-impulse response (FIR) filter in the time domain following the procedures outlined in \citet{Parsons2016}, which was applied to PAPER power spectrum analyses for crosstalk suppression \citep{Ali2015, Cheng2018, Kolopanis2019}.
In comparing this against the technique from this work, we use two different methods for low-pass filtering the output $\mathbf{T}$ modes, the first being a fringe-rate domain deconvolution (DEC), and the second being a Gaussian Process regression (GPR) with a fixed length-scale hyperparameter.
In summary, we find that the SVD-based algorithm with a GPR filtering provides the best systematic suppression outside the foreground wedge.
In addition, we find that all algorithms can be tuned to be safe against EoR signal loss in that the structure they subtract from the data can be confined to low fringe-rate modes where the EoR signal is subdominant.

Filtering visibilities by constructing an FIR filter that zeroes out low-fringe rate modes of the data is a technique used by the PAPER collaboration for suppressing slowly fringing systematics like crosstalk \citep[see Figure 10 of][]{Cheng2018}, and, given the convolution theorem, is identical to direct Fourier filtering.
In practice, this technique suffers from boundary effects, where because the visibility is not both infinitely sampled and periodic, the systematic subtraction closer to the starting and ending time boundaries is degraded.
Another approach is to use a deconvolution in the Fourier domain, which builds a model for the signal by employing the standard CLEAN algorithm \citep{Hogbom1974}.
By limiting the Fourier modes of the model components to low fringe-rate modes, one can construct a model that is by construction low-pass filtered.
This too suffers from time domain boundary effects, and in addition comes with all the subtleties of using the CLEAN algorithm to high dynamic range.

Modeling the time-covariance of the data with a Gaussian process (GP) eliminates the need to use a Fourier transform, because the modeling is done entirely in the time domain and does not assume exact periodicity of the underlying signal \citep{Rasmussen06}; although for our purposes it does assume the signal to be statistically stationary across time.
One major concern for a GP-based low-pass filter is the issue of what kinds of time structure it does and does not allow into the model.
In other words, we seek to understand its effective degrees of freedom while fitting the data.
Without going into too much detail, a Gaussian process models the covariance of the data through a kernel function, which sets how correlated two data points are given their distance from each other.
For example, a standard squared-exponential kernel describing the covariance between two points in time, $t_1$ and $t_2$, can be written as
\begin{align}
k(t_1, t_2|\ell) = \exp\left[-\tfrac{1}{2}(t_1-t_2)\ell^{-2}(t_1-t_2)\right],
\end{align}
where $\ell$ is a characteristic length scale of correlations and is a hyper-parameter of the kernel.
The GP maximum a posteriori (MAP) function (in addition to its credible intervals) can be calculated, and is generally the desired product of GP modeling \citep[Eqn. 2.23 of][]{Rasmussen06}.
By decreasing or increasing the length-scale hyper-parmeter, $\ell$, we can create GP best-fit functions that have more or less structure in them, respectively.

\begin{figure*}
\centering
\label{fig:gpr_noise_test}
\includegraphics[scale=0.6]{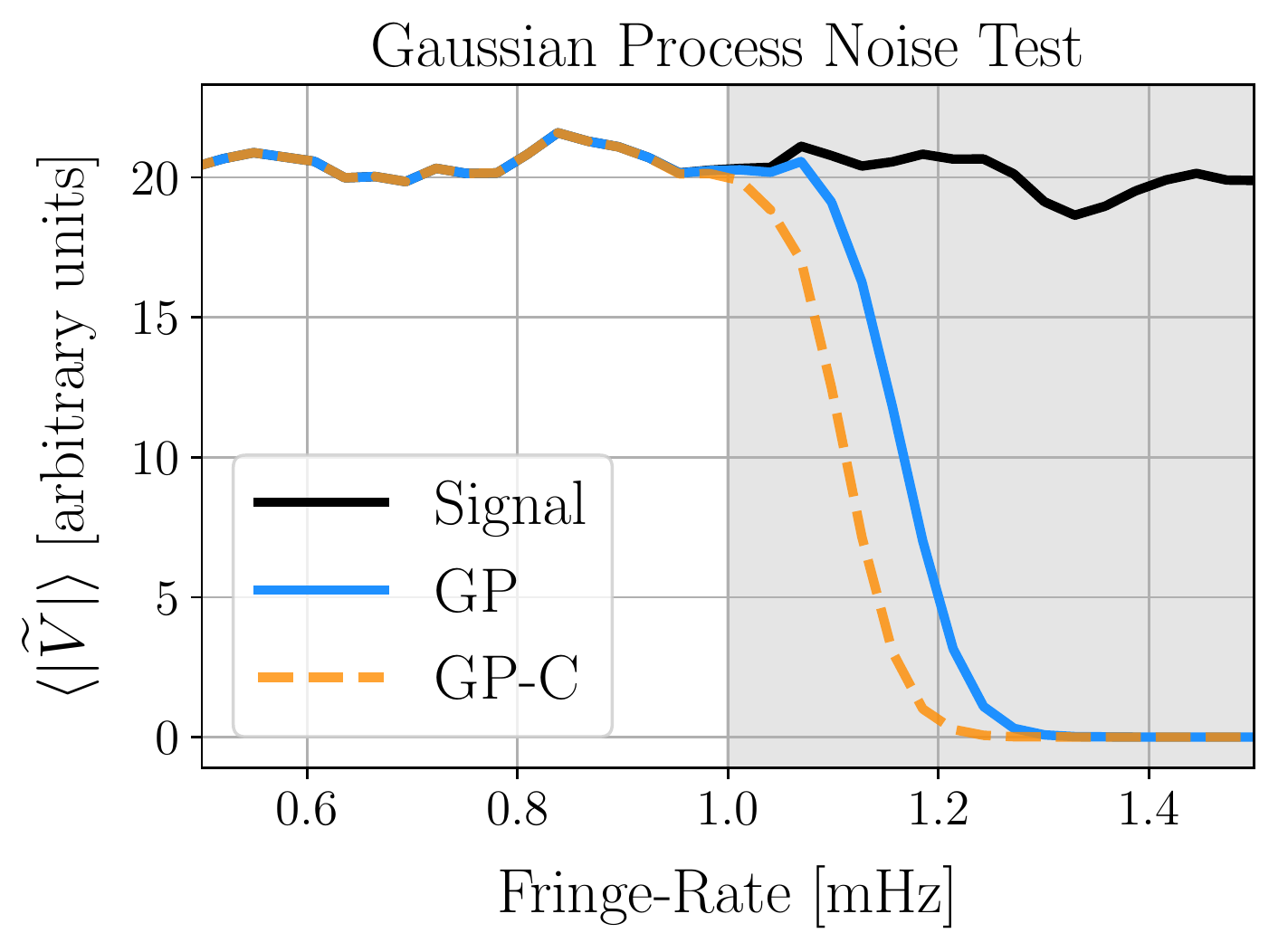}
\caption{Gaussian process fixed-hyper-parameter degree of freedom test, showing simulated noise (black), Gaussian process smoothing with a squared exponential kernel and a fixed length-scale hyper-parameter equivalent to $f_{\rm max} = 1.0$ mHz (GP) and constrained-Gaussian process smoothing with a fixed length-scale hyper-parameter equivalent to $f_{\rm max} = 0.95$ mHz (GP-C). We can see the former Gaussian process model (GP) picks up on signal power slightly outside of the fringe-rate cutoff defined by its length-scale hyper-parameter (shaded). Degrading this hyper-parameter by 5\% makes the model diverge from the signal at the desired fringe-rate cutoff (GP-C).}
\end{figure*}

\begin{figure*}
\centering
\label{fig:frfilter_comparison}
\includegraphics[scale=0.52]{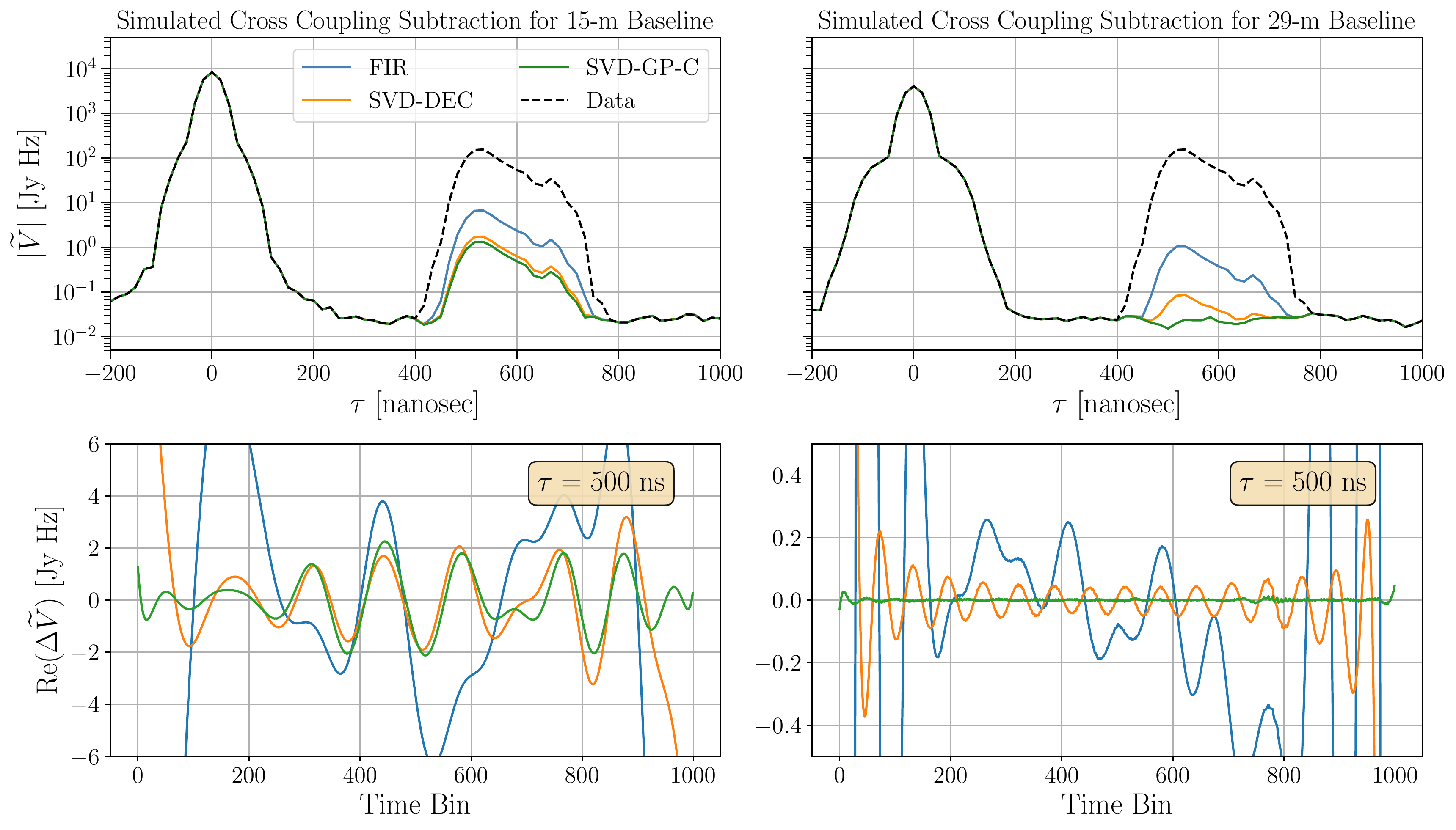}
\caption{Cross coupling removal comparison test for a finite-impulse response filter model (FIR), an SVD model with deconvolution low-pass filtering (SVD-DEC) and an SVD model with constrained Gaussian process low-pass filtering (SVD-GP-C), applied to a 15-meter baseline (left) and a 29-meter baseline (right). The top panels show modeling and removal in delay space, and the bottom panels show the difference of each method's systematic model with the true systematic as a function of time, highlighting the severe boundary effects suffered by the FIR and SVD-DEC techniques.}
\end{figure*}

To use a GP as a low-pass filter, we adopt a squared-exponential kernel with a fixed length scale hyper-parameter.
If the data is a time series, then this fixed length scale in time, $\ell$, sets the maximum fringe-rates allowed by the filter as $f_{\rm max} = \ell^{-1}$.
If we define the effective degrees of freedom of a model as the extent to which it allows power in fringe-rate space, then the effective degrees of freedom of a GP model can be tested via regressing over random noise simulations.
We generate 100 independent, uncorrelated complex Gaussian noise visibilities, compute our GP best-fit function given our fixed length-scale hyperparameter, and take its Fourier transform into fringe-rate space.
We then take the absolute value of each Fourier transformed GP fit and average over the independent realizations.
We do this for a GP with a $f_{\rm max} = 1.0$ mHz (GP) and for a GP with a more constrained cutoff of $f_{\rm max} = 0.95$ mHz (GP-C).
The results are shown in \autoref{fig:gpr_noise_test}, showing that indeed the normal GP allows power slightly beyond its cutoff scale, whereas the constrained version with a 5\% degraded cutoff diverges from the signal at the desired fringe-rate.
Therefore, when using a fixed-length-scale, squared-exponential GP as a low-pass filter, one should consider increasing the length-scale hyper-parameter by roughly 5\% its nominal value in order to ensure that the filter is not allowing power beyond the desired fringe-rate scale.

Next we compare how the several techniques outlined here perform on simulated visibilities containing only foregrounds and a cross coupling systematic.
Our comparison will focus on three methods for building a cross coupling visibility model: 1) a top-hat FIR filter that is 1 for all fringe-rate modes $|f| < f_{\rm max}$ and zero otherwise (labeled FIR), 2) the SVD-based method with a deconvolution low-pass filter applied to its $\mathbf{T}$ modes with a CLEAN boundary $|f| < f_{\rm max}$ (labeled SVD-DEC) and 3) the SVD-based method with a GP-C low-pass filter applied to its $\mathbf{T}$ modes (labeled SVD-GP-C).
The top panels in \autoref{fig:frfilter_comparison} plots the corrupted foreground + EoR + systematic visibility (black), with the systematic-subtracted visibility for each technique, for a 15-meter baseline (left) and a 29-meter baseline (right).
The bottom panels plot the difference of each systematic model with the true systematic at a single delay as a function of time.
We can clearly see the boundary effects that most severely plague the FIR method, but also the SVD-DEC method as well.
When the GP-C method brings the systematic down to EoR levels, it too suffers slightly from boundary effects, but they are significantly less prevalent than the boundary effects seen in the other techniques, highlighting one benefit of using a GP-based low-pass filter.

\section{Software}
\label{appendix:software}
The analysis presented in this work relies heavily on the Python programming language (\url{https://python.org}), and Python software developed by HERA collaboration members.
Here we provide a list of these packages and their version or Git hash:
\texttt{aipy [v2.1.12]} (\url{https://github.com/HERA-Team/aipy}), \texttt{healvis [v1.0.0]} \citep[\url{https://github.com/RadioAstronomySoftwareGroup/healvis};][]{Lanman2019}, \texttt{hera\_cal [v2.0]} (\url{https://github.com/HERA-Team/hera_cal}), \texttt{hera\_sim [v0.0.1]} (\url{https://github.com/HERA-Team/hera_sim}), \texttt{pyuvdata [v1.3.6]} \citep[\url{https://github.com/RadioAstronomySoftwareGroup/pyuvdata};][]{Hazelton2017}, and \texttt{uvtools [v0.1.0]} (\url{https://github.com/HERA-Team/uvtools}).
These packages in turn rely heavily on other publicly available software packages, including:
\texttt{astropy [v2.0.14]} \citep[\url{https://astropy.org};][]{Astropy2013}, \texttt{healpy [v1.12.9]} (\url{https://github.com/healpy/healpy}), \texttt{h5py [v2.8.0]} (\url{https://www.h5py.org/}), \texttt{matplotlib [v2.2.4]} (\url{https://matplotlib.org}), \texttt{numpy [v1.16.2]} (\url{https://www.numpy.org}), \texttt{scipy [v1.2.1]} (\url{https://scipy.org}), and \texttt{scikit-learn [v0.19.2]} (\url{https://scikit-learn.org}).

\bibliographystyle{apj}
\bibliography{systematic_modeling}

\end{document}